\documentclass[usenatbib]{mn2e}
\usepackage{amsmath, amssymb,color}
\usepackage{amsfonts}
\usepackage{epsfig,rotating}

\def\simeq{
\mathrel{\raise.3ex\hbox{$\sim$}\mkern-14mu\lower0.4ex\hbox{$-$}}
}

\def\ltsima{$\; \buildrel < \over \sim \;$}
\def\simlt{\lower.5ex\hbox{\ltsima}}
\def\gtsima{$\; \buildrel > \over \sim \;$}
\def\simgt{\lower.5ex\hbox{\gtsima}}

\def\msun{{\rm M_{\odot}}}

\def\be{\begin{equation}}
\def\ee{\end{equation}}
\def\le{{L_{\rm Edd}}}

\def\mbh{{M_{\rm BH}}}

\def\del#1{{}}

\newcommand{\apj}{ApJ}

\newcommand{\mnras}{MNRAS}
\newcommand{\aap}{A\&A}
\newcommand{\araa}{ARA\&A}
\newcommand{\apjl}{ApJL}
\newcommand{\aj}{AJ}
\newcommand{\nat}{Nature}

\title{Energy- and momentum-conserving AGN feedback outflows}
\author[Kastytis~Zubovas, Sergei~Nayakshin]{Kastytis~Zubovas$^1$ and Sergei~Nayakshin$^2$\\ 
$^1$ Center for Physical Sciences and Technology, Savanori\c{u} 231, Vilnius LT-02300, Lithuania\\
$^2$ Department of Physics and Astronomy, University of Leicester, University Road, LE1 7RH, Leicester, UK\\ 
{E-mail:~} {\rm kastytis.zubovas@ftmc.lt}}

\begin{document}

\maketitle

\begin{abstract}

It is usually assumed that outflows from luminous AGN are either in the
energy-conserving (non-radiative) or in the momentum-conserving (radiative)
regime. We show that in a non-spherical geometry the effects of both regimes
may manifest at the same time, and that it is the momentum of the outflow that
sets the $\mbh-\sigma$ relation. Considering an initially elliptical
distribution of gas in the host galaxy, we show that a non-radiative outflow
opens up a wide ``escape route'' over the paths of least resistance. Most of
the outflow energy escapes in that direction. At the same time, in the
directions of higher resistance, the ambient gas is affected mainly by the
incident momentum from the outflow. Quenching SMBH growth requires quenching
gas delivery along the paths of highest resistance, and therefore, it is the
momentum of the outflow that limits the black hole growth. We present an
analytical argument showing that such energy-conserving feedback bubbles
driving leaky ambient shells will terminate SMBH growth once its mass reaches
roughly the $M_\sigma$ mass derived earlier by \cite{King2003ApJ} for
momentum-conserving AGN outflows. Our simulations also have potentially
important implications for observations of AGN jet feedback and starburst
galaxy feedback. The collimation of the wide angle AGN outflow away from the
symmetry plane, as found in our simulations, indicates that credit for work
done by such outflows may sometimes be mistakenly given to AGN jets or star
formation feedback since wide angle $v \sim 0.1 c$ outflows are harder to
observe and the phase when they are present may be short.

\end{abstract}

\begin{keywords}
{quasars:general --- galaxies:evolution --- accretion,
  accretion discs --- black hole physics}
\end{keywords}

\section{Introduction}

Over the past decade, there has been growing interest in and
understanding of the role of feedback from active galactic nuclei
(AGN) in the evolution of galaxies. Astronomers now generally agree
that outflows caused by AGN feedback can drive gas out of galaxies and
quench star formation in the host \citep{Silk1998A&A, Page2012Natur},
thus establishing the observed correlations between the supermassive
black hole (SMBH) mass and host galaxy spheroid velocity dispersion
\citep{Magorrian1998AJ, Ferrarese2000ApJ, Tremaine2002ApJ,
  Gultekin2009ApJ}, dynamical mass \citep{Haering2004ApJ,
  McConnell2011Natur} and other parameters
\citep{Aller2007ApJ,Feoli2009ApJ}.

A considerable effort was extended to explain these findings from a
theoretical standpoint, but the problem remains far from being
solved. Both semi-analytical \citep{Croton2006MNRAS, Bower2006MNRAS}
and numerical simulations \citep[e.g.,][]{DiMatteo2005Natur} showed
that AGN feedback must be a key ingredient in galaxy formation and
evolution, especially at the high mass end. However, the precise
mechanism of the feedback communication from the AGN to galaxy's gas
remains elusive to this day. The situation is complicated since there
is no agreement on which process -- wide angle gas outflow, jet, or
radiation -- is the main mechanism of feedback, and also whether this
delivers energy (heating), physical push (momentum), or both, to the
ambient gas. For example, \cite{DiMatteo2005Natur, DiMatteo2008ApJ,
  Booth2009MNRAS} show that depositing $\sim 5 \%$ of the AGN
luminosity into the ambient gas during the rapid Eddington-limited
SMBH growth establishes the observed
correlations. \cite{Sijacki2007MNRAS} adds to the picture jets in form
of hot bubbles emitted by AGN at lower accretion rates, while jets in
\cite{Dubois2012MNRASb} also transfer momentum to the ambient gas.
\cite{Sazonov2005MNRAS} and \cite{Ciotti2007ApJ} propose that Compton
radiative heating of ambient gas by AGN radiation field plays a
significant role in limiting SMBH
growth. \cite{Fabian1999MNRAS,Thompson2005ApJ,DebuhrEtal11} suggest
that radiation pressure on dust is the main culprit of AGN feedback.
Wide angle outflows from AGN that deliver momentum to the ambient gas
are investigated by \cite{Debuhr2010MNRAS}. \cite{King2003ApJ,
  King2005ApJ} considers effects of a wide angle outflow on the
ambient gas; both momentum and energy of the outflow are important.

In this paper we focus on the effects of fast wide angle outflows from
AGN on the host galaxy gas in the context of the \cite{King2003ApJ}
model. Our main results are however more general, and add to a growing
body of work showing that the efficiency of energy deposition into the
ambient gas is actually quite low {\em if the gas is clumpy or
  inhomogeneously distributed}. \cite{WagnerEtal12} studied
numerically (using a grid-based code) the interaction of a powerful
jet with two-phase medium in the host galaxy, and found that the
efficiency of energy transfer to the cold medium is only $\sim 10$\%
(undoubtedly this particular number depends on the parameters of the
cold phase and, perhaps, numerical resolution). \cite{WagnerEtal13}
extended this work to the case of wide angle outflows, and found
similar results. These authors found that ``the outflow floods through
the intercloud channels, sweeps up the hot ISM, and ablates and
disperses the dense clouds. The momentum of the UFO is primarily
transferred to the dense clouds via the ram pressure in the channel
flow, and the wind-blown bubble evolves in the energy-driven regime.''

Bourne, Nayakshin \& Hobbs (2013; submitted, BNH13 hereafter) used a
completely different numerical technique -- smoothed particle
hydrodynamics \citep[SPH; see, e.g.,][]{Springel2010ARA&A}, employing
the ``SPHS'' algorithm of \cite{HobbsEtal13a} that is designed to
reduce artifical numerical effects of the classical SPH in a clumpy
medium. BNH13 obtained results very similar to that of
\cite{WagnerEtal13}, and proposed that this inefficiency of AGN
feedback energy deposition into the ambient medium explains how SMBH
can grow to the substantial masses observed despite producing huge
amounts of energy in the fast outflows that could destroy bulges of
host galaxies multiple times over. \cite{Nayakshin13b} included these
effects into an analytical study of AGN feedback and showed that the
observed $M-\sigma$ relation can be reproduced by such
energy-conserving flows {\em if star formation in clumpy medium is
  also taken into account.}

In this paper we perform numerical simulations with a ``classical''
SPH code that has a different implementation of AGN feedback compared
to either \cite{WagnerEtal13} or BNH13, and different initial
conditions for the ambient gas in the galaxy. We consider initially
homogeneous, rather than clumpy, medium but distribute it in a
non-spherical geometry. As an example, we consider AGN feedback on an
elliptically distributed ambient gas in a galaxy, so that gas density
along the galactic plane is highest and drops gradually to the lowest
value perpendicular to the plane. This geometry should be considered
as a simplest rudimentary step closer to realistic galaxies, which are
mostly non-spherical except perhaps in the case of ``red-and-dead''
elliptical galaxies.

Despite these numerical and set-up differences with previous studies,
we recover the main conclusions of \cite{WagnerEtal13} and BNH13. We
find that the AGN feedback quickly inflates two outflow bubbles
perpendicularly to the galactic plane, where the gas density is
lowest. Most of the feedback energy escapes through these funnels,
leaving the denser gas exposed mainly to the momentum of the AGN
wind. Therefore, the dense gas behaves as if it were affected by
momentum feedback only. The energy-momentum separation found in our
simulations is large-scale rather than local, small scale, as in
\cite{WagnerEtal13} and BNH13, but the final conclusions are similar.

We also provide a simple analytical argument \citep[related to an
  earlier study of ``leaky feedback bubbles'' in the content of
  stellar feedback by][]{HMurray09} that confirms the main result of
\cite{Nayakshin13b}: since the cold gas is momentum-driven, the SMBH
mass required to expel the cold shell in this {\em energy-conserving}
regime is similar to the momentum-driven result of \cite{King2003ApJ},
which in itself is pleasingly close to the observed $M-\sigma$
relation \citep{Ferrarese2000ApJ, Tremaine2002ApJ, Gultekin2009ApJ}.

The paper is structured as follows. We begin with a brief review of
the state-of-the-art of our understanding of the physics of ultra-fast
outflows in spherically symmetric models in \S \ref{sec:radiation} -
\ref{sec:ufos}. We then present in \S \ref{sec:toy} a toy model
spherically symmetric ``leaky shell'' calculation that takes into
account energy escape from the bubble via low density channels in the
ambient shell.  In Section \ref{sec:nummodel} we describe the setup of
numerical simulations and in Section \ref{sec:results} we present
their results. We follow with a discussion in Section
\ref{sec:discuss} and summarize and conclude in Section
\ref{sec:concl}.

\section{Theoretical Preliminaries} \label{sec:largescale}

\subsection{Radiation from AGN}\label{sec:radiation}

Direct AGN radiation impact on gas in the host galaxy is unlikely to
be the main driver of SMBH-galaxy coevolution. Radiation pressure
effect on dust even in initially homogeneous ambient medium is likely
to be limited to momentum push of $\sim L_{\rm Edd}/c$ due to
development of radiation Rayleigh-Taylor instability
\citep{KrumholzThompson13}. This then falls short by a factor of at
least $\sim 10$ of what is required to drive the gas out of the host
completely \citep{SilkNusser10}. Furthermore, it is fairly obvious
that effective radiation pressure on cold clumps in a clumpy
multi-phase medium would be even less significant than in the
initially uniform ambient gas shell, since most of the AGN radiation
field may never impact the cold clouds if their covering angle as seen
from the AGN is small. In contrast to \cite{Murray2005ApJ}, we
therefore doubt that radiation pressure is ever enough to put SMBH on
their $M-\sigma$ relations.

Even for homogeneous gas distributions, radiation heating by Compton
effect is likely to be efficient only at low gas densities, that is,
for galaxies at low redshift \citep[see][]{Sazonov2005MNRAS}, whereas
most of SMBH growth occurs at high redshift. For inhomogeneous, i.e.,
clumpy multi-phase medium, the rate of radiative heating of cold gas
becomes too low to heat it to high temperatures
\citep{Sazonov2005MNRAS}. Furthermore, the column density of denser
cold clumps can be much greater than 1 g cm$^{-2}$ (cf. BNH13), so
that soft and medium energy X-rays do not penetrate the clouds. This
further reduces the significance of AGN radiative input into the gas.

\subsection{Relativistic jets from AGN}\label{sec:jets}

Relativistic jets emanating from AGN are implicated as important
feedback sources, especially in galaxy cluster environment
\citep[see][for a recent review]{Fabian2012ARA&A}. Here we concentrate
on the early gas-rich epoch of isolated galaxies, where AGN are likely
to be in the ``quasar mode'' when jets are not expected to be
crucially important
\citep[e.g.,][]{ChurazovEtal05,Sijacki2007MNRAS}. In addition, as
\cite{WagnerEtal13} demonstrate, the effects of wide-angle and
collimated outflows in the energy-conserving regime (negligible
radiative cooling) are actually very similar after the outflow shocks
against the ambient medium, which always inflates a wide angle hot gas
bubble. Therefore we only study wide-angle outflows below, but expect
the main results to be similar for jets in the energy-conserving
regime.

\subsection{Ultra-fast wide angle outflows}\label{sec:ufos}

The AGN wind feedback model of
\cite{King2003ApJ,King2005ApJ,King2010MNRASa} is very attractive for a
number of reasons. First of all, it is based on observations
\citep[e.g.,][]{Pounds2003MNRASa, Pounds2003MNRASb,
  Tombesi2010A&A,PoundsVaughan11, Pounds2011MNRAS} that show that
luminous AGN \citep[$L_{\rm AGN} \ga 0.01 L_{\rm
    Edd}$,][]{King2013ApJ} drive powerful and fast ($v_{\rm w} \sim
0.1 c$) winds. These observational results are deeply natural on
theoretical grounds \citep{King2003MNRASb}, first pointed out in the
classical ``Standard Accretion Disc Theory'' of \cite{Shakura1973A&A}:
AGN radiation escaping to infinity is likely to accelerate gas to
velocity of order the local escape velocity
\citep[e.g.,][]{Proga2000ApJ, Everett2004ApJ}, which is as high as
$\sim (0.1-0.3) c$ in the innermost region of accretion discs, where
most AGN radiation is produced. For SMBH accreting at near Eddington
accretion rates the momentum flux is expected to be $\sim L_{\rm
  Edd}/c$.

An important element of the physics of the model is the presence of
two modes of feedback. One, called the momentum-driven outflow, is
appropriate for the case when the relativistic wind emanating from the
vicinity of the SMBH shocks against the ambient medium and cools
efficiently, transferring only its ram pressure (and, hence, momentum
flux) to the ISM. This mode of feedback is essential in explaining the
M-sigma relation \citep{King2010MNRASa}. Similar arguments based on
energy-conserving outflows predict SMBH masses that are several orders
of magnitude lower than observed\footnote{unless an arbitrary and
  significant reduction in feedback efficiency is introduced.}
\citep{Silk1998A&A,King2010MNRASb}. The second type of AGN wind
feedback, termed the energy-driven or the energy-conserving outflow,
occurs when the cooling of the shocked primary outflow from the SMBH
is inefficient and leads to a hot wind bubble expanding from the
centre of the galaxy, transferring most of its kinetic luminosity
$L_{\rm kin} \simeq 0.05 L_{\rm AGN}$ to the ISM. This type of outflow
can drive {all} of the ambient gas of the host galaxy out at
velocities exceeding $1000$~km/s and drive mass outflows of several
thousand $\msun$~yr$^{-1}$ \citep{King2011MNRAS,Zubovas2012ApJ}, in
accordance with the recently-observed molecular outflows in several
active galaxies \citep{Feruglio2010A&A,Rupke2011ApJ,Sturm2011ApJ}.
The model may also account for the ``Fermi bubbles'' in the Milky Way
\citep{Zubovas2012MNRASa}.

At a fixed SMBH mass and luminosity, the regime in which the UFO
impacts the gas in the host depends mainly on the location of the
contact discontinuity between the wind driving the shock and the
outflowing ISM. If the contact discontinuity is closer than some
critical distance $R_{\rm C}$ (called the cooling radius) from the
AGN, the shocked wind cools efficiently via inverse-Compton scattering
\citep{Ciotti1997ApJ,King2003ApJ}, resulting in a momentum-driven flow
\citep{King2010MNRASa}. When the contact discontinuity moves beyond
$R_{\rm C}$, the shocked wind no longer cools, leading to an
energy-driven outflow \citep{King2011MNRAS}. The velocity of this
outflow propagating in an isothermal potential with velocity
dispersion $\sigma \equiv 200 \sigma_{200}$~km/s is
\citep[cf.][]{King2005ApJ,King2011MNRAS}
\begin{equation} \label{eq:ve}
v_{\rm e} = \left(\frac{2 \eta \sigma^2 c}{3} \frac{f_{\rm c}}{f_{\rm
    g}}\right)^{1/3} = 925 \sigma_{200}^{2/3} f^{-1/3} {\rm km
  s}^{-1},
\end{equation}
where $\eta \simeq 0.1$ is the radiative efficiency of accretion, $c$
is the speed of light, $f_{\rm g} \equiv \rho_{\rm g} / \rho_{\rm
  tot}$ is the gas fraction (assumed constant with radius) and $f_{\rm
  c} = 0.16$ is its cosmological value. In the second equality, we
scaled the result to $\sigma_{200}$ and $f \equiv f_{\rm g}/f_{\rm
  c}$. \citet{Zubovas2012MNRASb} calculate the cooling radius to be
$R_{\rm C} \sim 500$~pc for typical AGN parameters; this large value
suggests that momentum-driven outflows are important in the central
parts of galactic bulges and are the relevant mode of feedback for
establishing the $M-\sigma$ relation.

However, the derivation of this cooling radius depends on the
assumption that the shocked wind contains ions and electrons with the
same temperature $T_{\rm i} \sim 10^{11}$~K (hence we call this the
``One temperature'', or 1T, model). \citet{Faucher2012MNRASb} showed
that when the energy equilibration timescale is accounted for, the
electrons in the shocked wind only reach temperatures of $T_{\rm e}
\sim 3 \times 10^9$~K. The lower than expected $T_{\rm e}$ may be the
reason why the Inverse Compton (IC) radiation from the cooling reverse
shock of the fast outflows from AGN has not yet been observationally
identified \citep[][BN13 hereafter]{Bourne2013arXiv}. Potential
implications of an inefficient electron-ion coupling are very
significant for this AGN feedback model.  At $T_{\rm e}\simlt 3\times
10^9$~K, the cooling rate via IC scattering on the AGN radiation field
- the primary cooling mechanism acting on the wind - is much smaller
than in the 1T model, leading to a situation where the outflow is
energy-driven at every distance from the AGN. Even within the 1T
model, properly accounting for the accumulation of energy in the
shocked wind leads to a much lower value of the derived cooling radius
\citep{McQuillin2013MNRAS}, so that AGN outflows should generally be
in the energy-driven regime.

\subsection{A toy model: a leaky shell}\label{sec:toy}

At first glance, the results of \citet{Faucher2012MNRASb} and BNH13
are very troubling for the wind feedback model. If the outflows are
energy-driven even at the very centre of the galaxy, then the SMBH
needs a much lower mass in order to expel the surrounding gas, quench
its own fuel supply and thus establish the M-sigma relation. However,
Nayakshin (2013; N13 hereafter) noted that {\bf the feedback model of
  \citet{King2003ApJ}} also assumes that the ISM in the galaxy is
spherically symmetric and smooth. As a result, the AGN expels gas
evenly in all directions and the fuel supply is cut once the outflow
bubble begins expanding. Arguing instead that the ISM of the host
galaxy's bulge is clumpy, and that the clumps are over-taken by the
outflow easily, N13 showed that it is only the momentum of the AGN
outflow that matters in expelling the high density clumps out.
Further, the densest clumps are self-gravitating and will form stars
sooner than they could feed the SMBH. Using these two constraints
together, N13 showed that the SMBH stops growing when it reaches the
mass of order the $M_\sigma$ mass derived by \cite{King2003ApJ},
despite the outflow being in the energy-driven regime.

We now use a toy quasi-spherical model to argue that energy-conserving
AGN outflows acting on broken up ``leaky shells'' would push them
mainly by the ram pressure of the outflow, and that this fact actually
leads to an $M-\sigma$ relation that is quite similar to that derived
by \cite{King2003ApJ}. Our conclusions here are similar to those
reached by N13.

Following \cite{King2003ApJ}, we write down the momentum equation for
a swept-up shell of ambient gas at radius $R$. In the singular
isothermal potential (SIS) with 1D velocity dispersion $\sigma$, the
shell's mass is $M_{\rm g}(R) = 2 f_{\rm g} \sigma^2 R/G$. The
momentum equation for the shell moving outward with velocity $v = \dot
R$ is \citep{King2005ApJ}
\begin{equation}
\frac{d}{dt}\left[M_{\rm g}\left(R\right) \dot R\right]= -\frac{G
  M(R)M_{\rm g}(R)}{R^2} + 4\pi R^2 P\;,
\label{mom1}
\end{equation}
where $M(R) = M_{\rm g} f_{\rm g}^{-1}$ is the total (dark matter,
stars and gas) enclosed mass inside radius $R$.  Note that the term $G
M(R)M_{\rm g}(R)/R^2 = 4 f_{\rm g} \sigma^4/G$ for the SIS potential,
so this can be simplified to
\begin{equation}
\frac{d}{dt}\left[M_{\rm g}\left(R\right) \dot R\right]= - \frac{4
  f_{\rm g} \sigma^4}{G}+ 4\pi R^2 P\;.
\label{mom2}
\end{equation}
In the momentum-driven limit, \cite{King2003ApJ} derived a theoretical
$M-\sigma$ relation,
\begin{equation}
\mbh = M_\sigma = f_{\rm g} \frac{\kappa \sigma^4}{\pi G^2}\approx 3.78
\times 10^8 \msun\; f \; \sigma_{200}^4\;.
\label{msigK}
\end{equation}
where $\kappa$ is the electron scattering opacity. Equation
(\ref{msigK}) has no free parameter except for $f_{\rm g}$. In young
gas-dominated galaxies, this parameter is close to one. Later, star
formation depletes the gas, while larger scale processes, such as
cooling flows from cluster environments, replenish its
content. \citet{Zubovas2012MNRASb} showed that these processes can
produce a spread in the final SMBH masses of a factor $\sim 4$. If one
assumes that $f_{\rm g} \sim 0.16$, eq. (\ref{msigK}) gives a result
within a factor of $\sim 2$ of the observed $M-\sigma$ correlation
\citep{Gultekin2009ApJ,Ferrarese2000ApJ}. In reality, this assumption
is not quite correct, because the gas fraction in galaxies varies with
mass and star formation rate \citep{Daddi2010ApJ,Santini2014A&A} and
large gas fractions are expected in galaxies in the early Universe
\citep{Dubois2012MNRAS}. In fact, a systematically varying gas
fraction might explain why the observed $M-\sigma$ relation has
steeper slope than predicted by eq. (\ref{msigK})
\citep{Zubovas2012MNRASb}.

We now consider the energy driven limit.  Since the bubble expands
very slowly compared with the sound speed of the hot gas filling it,
$c_{\rm b}\sim v_{\rm w}$, one may expect that the pressure, density
and temperature within the bubble will be approximately uniform.  Let
$E = (4\pi R^3/3) (3/2 P) = 2\pi R^3 P$ be the bubble's thermal
energy; here we assumed that adiabatic index is $\gamma=5/3$. The
energy equation for the bubble is therefore \citep{King2005ApJ}
\begin{equation}
\frac{d}{dt} \left[4\pi R^3 P\right] = \frac{\eta L}{2} - P
\frac{dV}{dt} - \frac{4 f_{\rm g} \sigma^4}{G}\dot R - \frac{2\pi R^3
  P}{t_{\rm esc}}\;.
\label{en1}
\end{equation}
On the right hand side of this equation, the first term is the energy
deposition rate into the bubble due to the wind launched by an AGN
with luminosity $L = lL_{\rm Edd}$ ($L_{\rm Edd}$ being the Eddington
luminosity), the second term is the adiabatic expansion losses for the
bubble, with $V= 4\pi R^3/3$ being the bubble's volume, the third one
is the work done against gravity when lifting the shell out of the
potential. Finally, the last term is new compared to
\cite{King2005ApJ} and is a parameterisation of the energy escape from
the bubble. Expressed as $E/t_{\rm esc}$, it is a generic loss term,
applicable to all losses where the loss timescale does not depend
strongly on bubble size, shape or energy content. In this case, we
consider adiabatic energy escape with gas leaking out of the
bubble. We reason that if there are escape channels for the hot gas,
the gas escapes moving at about the sound speed of the hot gas,
$c_{\rm b}$. The rate of the mass flow through the opening is
approximately $\rho_{\rm b} \Omega R^2 c_{\rm b}$, where $\Omega$ is
the solid angle of the escape route. The rate of energy loss rate from
the bubble through the channel is $\sim \Omega R^2 c_{\rm b} (E/V) =
\Omega R^2 c_{\rm b} (3/2) P \equiv 2\pi R^3 P/t_{\rm esc}$. The last
equation defines $t_{\rm esc}$ as
\begin{equation}
t_{\rm esc} = \frac{R}{c_{\rm b}} \; \frac{4\pi}{3\Omega} \equiv
\lambda \frac{R}{c_{\rm b}}\;.
\label{tesc1}
\end{equation}
The escape time is therefore roughly the sound crossing time of the bubble
divided by the fraction of the angle open for direct escape.

As in \cite{King2011MNRAS}, we seek a constant velocity solution $R =
v t$ for the expanding bubble. Equation (\ref{mom2}) is then
\begin{equation}
4\pi R^2 P = \frac{2 f_{\rm g} \sigma^2}{G}\left( v^2 + 2
\sigma^2\right)\;.
\label{peq1}
\end{equation}
Substituting this into equation (\ref{en1}), and using equation
(\ref{tesc1}), we obtain a cubic equation for outflow velocity,
\begin{equation}
3 v^3 + 10 \sigma^2 v + \frac{c_{\rm b}}{\lambda}\left(v^2 +
2\sigma^2\right) = \frac{\eta G L}{2 f_{\rm g} \sigma^2}
\label{v3a}
\end{equation}
This equation is identical to that obtained by \cite{King2011MNRAS}
except for the third term on the left side of the equation. This term
is responsible for energy escape from the bubble. In the limit of
negligible energy escape from the bubble, $\lambda\rightarrow \infty$,
we recover the approximate solution of \cite{King2011MNRAS}, $v
\rightarrow v_{\rm e}$, as given by equation (\ref{eq:ve}).

In this section we are especially interested in the opposite limit,
when the reverse shock energy leaks out of the shell rapidly, that is
$\lambda \sim$ a few. In this case the third term on the left of
equation (\ref{v3a}) is the dominant one. Omitting all the other terms
on the left, and setting $c_{\rm b}\approx \eta c$, we obtain
\begin{equation}
v^2 \approx 2 \sigma^2 \left(l \lambda \frac{\mbh}{M_\sigma} -1\right)\;.
\label{limit2}
\end{equation}
We see that in this case, provided $l\sim 1$, $v\approx \sigma$, i.e.,
the expansion of the bubble is much slower. This makes perfect
physical sense: since the energy of the hot bubble can make its way
out of the bubble through empty channels, the bubble pressure and
energy are much smaller than they are in the opposite case, and so the
expansion is slower.

Further, if we set $l=1$ (that is, $L = \le$), we see that to drive
the shell outward at all,
\begin{equation}
\mbh \ge \lambda^{-1} M_\sigma\;,
\label{msigma_alt}
\end{equation}
which shows that the SMBH mass should be near the $M_\sigma$ mass
\citep{King2003ApJ} for an {\em energy-conserving} outflow to drive
the shell outward if the bubble is strongly leaky, e.g.,
$\lambda\sim$~a few.

Finally, substituting the approximate solution (\ref{limit2}) into
equation (\ref{peq1}), we find that the force applied by the bubble
onto the ambient shell is
\begin{equation}
4\pi R^2 P \approx \frac{\lambda l \le}{c} = \frac{\lambda L}{c} 
\label{peq2}
\end{equation}
This equation shows clearly that when energy leaking is significant,
i.e. $\lambda\simgt 1$, the force acting on the shell is only a
little larger than that due to the ram pressure of the UFO in the
momentum-conserving limit, $L/c$.

Our derivation shows that an energy-conserving (i.e. non-radiative)
bubble that has wide escape channels for hot gas drives the ambient
shell outwards with a much smaller efficiency, consistent within a
factor of order unity with the momentum-driven regime. The physical
origin of this result is transparent. Mechanical loss of bubble energy
through cavities is a form of cooling (adiabatic expansion rather than
radiative). The bubble is therefore less energetic than in 1D solution
without energy escape channels \citep{King2005ApJ}. The limiting case
of a large fraction $\lesssim 1$ of the outflow energy escaping
through the open channels should produce outward pressure comparable
with just the ram pressure of the wind; hence the result.

This derivation may alleviate a certain tension between the AGN wind
feedback model of \citet{King2003ApJ} and numerical simulations of
galaxy formation which include AGN feedback
\citep[e.g.,][]{DiMatteo2005Natur, Sijacki2007MNRAS, Booth2009MNRAS,
  Dubois2012MNRAS}. These simulations typically adopt a coupling
factor $\epsilon_{\rm f} \simeq 5-15 \%$ between the AGN luminosity
and kinetic outflow energy; the parameter is calibrated in order for
the simulations to produce SMBH masses in accordance with
observations. This is consistent with the amount of energy injected by
the fast outflow into the host galaxy in the \citet{King2003ApJ}
model. However, in the latter model most of this energy is lost to IC
radiation within the cooling radius, so that the outflow is
momentum-driven, whereas the simulations cited above do not include
the semi-relativistic IC energy losses and should thus be
energy-conserving instead.

It is then somewhat puzzling why the simulations recover the correct
M-sigma relation despite assuming a different physical mechanism. A
possible solution to this issue is that the cosmological simulations
cited above may well contain the "leaky shell" effects considered in
this section. If this is the case, then, although the feedback energy
injected into the gas is much higher than necessary for gas expulsion,
most of this energy escapes through low-density gaps in the gas
distribution. The momentum given to cold dense gas is then much lower
than in spherically symmetric analytical models, explaining the
discrepancy.

Our toy analytical model makes several simplifying assumptions
regarding the structure of the galaxy. One such assumption is the
isothermal density profile of the background potential. This
assumption is roughly correct for early-type galaxies
\citep{Koopmans2009ApJ}, but in general, galaxies have somewhat
triaxial dark matter distributions
\citep{Bailin2005ApJ,Diemand2011ASL}. In the cases where the
background potential density is higher in the plane of the galaxy, the
effect we describe becomes even more pronounced, as the gas has an
easier time escaping in the polar direction. In the opposite case,
where the dark matter halo density is higher in the polar direction,
the effect is somewhat mitigated, but not negated completely, since in
the central parts of the galaxy, baryonic matter dominates the
potential and its distribution is more important that the variations
in dark matter density.

\section{Numerical model}\label{sec:nummodel}

We now design a numerical experiment to see if energy can indeed
``leak'' in non-spherical geometries.  Unlike BNH13, we run SPH
simulations of energy-driven AGN feedback on homogeneous surrounding
gas distributions. We use the hybrid N-body/SPH code Gadget-3
\citep[an updated version of the code by][]{Springel2005MNRAS}. The
code employs adaptive smoothing lengths for both hydrodynamics and
gravity \citep[see][for a review of SPH methods in
  astrophysics]{Springel2010ARA&A} in order to keep the same number of
neighbours within a particle's smoothing kernel. We utilize a standard
cubic spline kernel with 40 neighbours.

Our simulations contain gas and the SMBH embedded in a static
isothermal background potential with $\sigma = 200$~km/s. The gas is
distributed in a spherical shell with $R_{\rm out} = 5$~kpc and
$R_{\rm in} = 200$~pc, and the total mass of the gas is set at $M_{\rm
  gas} = 0.16M_{\rm pot}\left(<5{\rm kpc}\right) \simeq 1.5 \times
10^{10} \; \msun$. Each simulation uses $10^6$ particles, giving a
particle mass of $m_{\rm SPH} = 1.5 \times 10^4 \; \msun$ and mass
resolution $m_{\rm res} = 40 m_{\rm SPH} = 6 \times 10^5 \; \msun$. We
take the SMBH mass to be the formal critical mass that allows driving
the gas out by pure momentum feedback: $M = M_\sigma = 3.68 \times
10^8 \; \msun$.

The gas is distributed in a way to give higher density in the
equatorial plane and lower density at the poles. Numerically, we do
this by a transformation tan$\theta \rightarrow \zeta$tan$\theta$,
where tan$\theta = |z|/r_{\rm cyl}$ and $\zeta$ is a free
parameter. We investigate three cases: $\zeta = 1$ (spherically
symmetric), $\zeta = 2$ and $\zeta = 5$. For $\zeta = 2$, the
effective gas fraction in the initial distribution varies from $0.035$
in the polar direction and $0.33$ in the equator; for $\zeta = 5$, the
values are $0.006$ and $0.6$, respectively. In addition to this
``squeezing'', for each case we investigate the effect of rotation by
running a model with $v_{\rm rot} = 0$ and a model with $v_{\rm rot} =
\sigma$ around the Z axis; given the background potential and the gas
distribution, the circular rotation velocity in the midplane is
$v_{\rm circ} \simeq \sqrt{2\times 1.16}\sigma \simeq 1.5 \sigma$. We
label the six models by their $\zeta$ values and the presence of
rotation: Z0R0, Z0R1 and so on (see Table \ref{table:param}).

\begin{table}
\centering
\begin{tabular}{c | c | c | c }
$ $ & $\zeta = 1$ (spherical) & $\zeta = 2$ & $\zeta = 5$ \\
\hline
$v_{\rm rot} = 0$ & Z1R0 & Z2R0 & Z5R0 \\
\hline
$v_{\rm rot} = \sigma$ & Z1R1 & Z2R1 & Z5R1 \\
\hline
\end{tabular}
\caption{The six models analyzed in the paper. The parameter $\zeta$
  refers to the squeezing of gas toward the midplane, via the
  transformation tan$\theta \rightarrow \zeta$tan$\theta$, with
  $\theta$ being the polar angle of the gas position. $\zeta = 2$
  produces an initial density contrast between polar and equatorial
  directions of $1:10$, while $\zeta = 5$ gives a contrast of
  $1:100$. Gas rotation is also investigated, but found to have very
  little effect on the results. See text for more information.}
\label{table:param}
\end{table}

BNH13 studied the response of the ambient inhomogeneous medium to a
single hot bubble inserted by hand in the initial condition of the
simulation. Here we implement continuous AGN feedback by using the
``virtual particle'' method \citep{Nayakshin2009MNRASb}. With this
approach, the AGN emits feedback particles that carry momentum and
energy. The momentum carried by a single virtual particle is chosen to
be $0.1 m_{\rm SPH} \sigma$, which is sufficiently small to preserve
simulation accuracy \cite[see][]{Nayakshin2009MNRASb}.

The number of particles emitted is then determined from the condition
that the total momentum emitted by the SMBH is $L_{\rm AGN} \Delta
t_{\rm BH} / c$, where $\Delta t_{\rm BH}$ is the current black hole
timestep. Typically, there are $\sim 3 \times 10^4$ virtual particles
in the simulation at any given time, although this number grows as the
volume swept up by the black hole outflow increases. While the above
number of virtual particles may seem small, one need to remember that
virtual particles are continuously created and destroyed (see
below). Since they move through the simulation volume at speeds much
exceeding the mean gas velocities, this implies that the total number
of virtual particles created and discarded during a simulation usually
significantly exceeds the total number of SPH particles.

Operationally, virtual particles propagate in straight lines radially
away from the SMBH with velocities $v_{\rm virt} = 0.1 c$ and interact
with SPH particles only when they are within a smoothing kernel of one
or more SPH particles. In this paper, both momentum and energy of
virtual particles is transferred to the SPH neighbour particles. The
rate at which different neighbour particles receive their energy and
momentum kicks is proportional to their contribution to the local gas
density (that is, the local value of their SPH kernel). This feature
of the code is important in properly distributing AGN feedback in a
multi-phase environment.

The virtual particle energy and momentum are reduced with each
interaction over a length of about one SPH smoothing lengh, to prevent
numerical artefacts, and the particles are removed from the simulation
when their momentum drops to negligible values. The timesteps of
virtual particles are carefully monitored so that they do not skip
interactions with dense (and hence compact) gas regions. With this
method, our results are thus complimentary to those obtained by BNH13
not only in terms of initial conditions but also in terms of numerical
methods.

The mass of the SMBH is assumed to stay constant (we track particle
accretion, but do not add their mass to the SMBH mass, because the gas
accretes on a viscous timescale $t_{\rm visc} \gg t_{\rm sim}$, the
duration of our simulations) and the AGN radiates at its Eddington
luminosity and emits virtual particles moving radially with a constant
velocity $v_{\rm w} = 0.1 c$, carrying a total momentum $\dot p =
L_{\rm Edd}/c$ and energy $\dot E = 0.05 L_{\rm Edd}$. When the
virtual particle enters an SPH particle's smoothing kernel, momentum
and energy is transferred to the SPH particle. The transfer happens
over several timesteps and several SPH particles can be affected by a
single virtual particle. In order to minimize the noise from
stochastic variations in virtual particle positions, we calculate
their number so that a single virtual particle has $p_{\rm virt} = 0.1
m_{\rm SPH} \sigma$. Virtual particles that lose $99\%$ of their
initial momentum (and energy, because the two are transferred in
exactly the same way) are removed from the simulation.

In addition to SMBH wind feedback, ambient gas is also affected by
radiative heating from AGN radiation. The gas is allowed to cool by
bremsstrahlung, IC and metal recombination line emission. We model
these processes and the radiative heating from the AGN by using the
heating-cooling prescription from \citet{Sazonov2005MNRAS}. These
authors calibrated their heating-cooling curve to a typical AGN
spectrum, which typically has $\sim20\%$ of the bolometric power in
X-rays. Their ionization parameter is defined with respect to the
total bolometric luminosity, and this is what we do as well. The
prescription assumes that all of the gas is optically thin to the
photoionizing radiation; given that the inner edge of our gas
distribution is at $200$~pc, the column depth is of order
$10^{24}$~cm$^{-2}$, i.e. the gas is marginally Compton-thin. In
denser regions, where the gas would be self-shielding, a more proper
treatment of gas optical depth would only enhance the results seen in
our simulations, i.e. cooling and fragmentation of gas into numerous
dense clumps. We modify the cooling function by suppressing gas
cooling at temperatures above $T_0 = 2 \times 10^8$~K with an extra
factor exp$(-T/T_0)$, in order to prevent numerical overcooling of gas
on the inner boundary of the shocked wind bubble. This is invoked
since the thickness of the layer in which the virtual particles
deposit their energy may be somewhat over-estimated, hence
under-predicitng the temperature of the shocked gas. Our main results
are very insensitive to this code detail, however.

In order to speed up the simulations, we convert gas particles into
star particles according to a Jeans' condition. Whenever a gas
particle density increases above the critical value
\begin{equation}
\rho_{\rm J} = \left(\frac{\pi k_{\rm B} T}{\mu m_{\rm p} G}\right)^3
m_{\rm sph}^{-2} \simeq 3 \times 10^{-16} T_4^3 \; {\rm g cm}^{-3}
\simeq 10^8 T_4^3 \; {\rm cm}^{-3},
\end{equation}
where $T_4 \equiv T/10^4$~K, it is converted into a star particle of
the same mass, which subsequently interacts with the other particles
only via gravity. The critical density assumes that the Jeans mass is
resolved in the gas. For the warm ISM with $T \simeq 10^4$~K, the
temperature floor of our model, the typical densities are $10^1-10^2
\; {\rm cm}^{-3} \ll \rho_{\rm J}$, so we are confident that we do not
overpredict the formation rate of gravitationally bound clumps. We
choose a Jeans' criterion, instead of a simple temperature-independent
density threshold, for the formation of sink particles in order to
ensure that there is no spurious sink particle formation at shock
fronts.

\section{Results}\label{sec:results}

The simulation results are presented below. We focus on three main
aspects of model evolution: the morphology of outflowing gas, the
distribution of gas velocities in polar and equatorial directions and
the distribution of energies for particles of different
density. Together, these indicators reveal that dense gas is pushed
away almost exclusively by the momentum input from the AGN, while the
diffuse gas carries away most of the wind energy.

\subsection{Gas morphology}

\begin{figure}
  \centering
    \includegraphics[width=0.45 \textwidth]{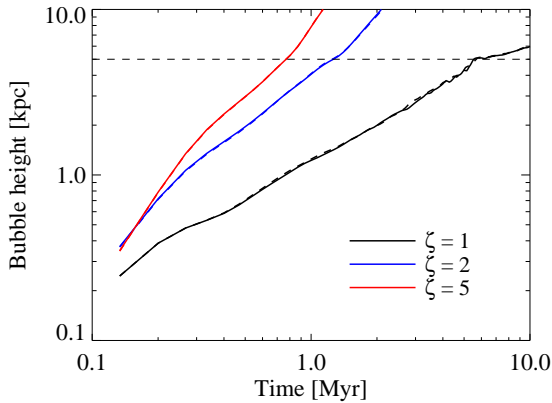}
  \caption{Height of the bubble in the polar direction
    ($|z|/\sqrt{x^2+y^2} > 4$). Black lines depict Z1 (spherical
    models), blue lines Z2, red lines Z5; dashed lines show rotating
    models (R1), solid lines show initially static ones (R0). The
    lines were smoothed over five data points with a weighted kernel
    to reduce numerical noise. A clear dichotomy between
    spherically-symmetric and anisotropic initial conditions is
    visible, with the diffuse gas in the Z2 and Z5 simulations blown
    away very quickly, reaching $h = 5$~kpc, i.e. the outer edge of
    the initial gas distribution, in $\sim 1.3$ and $\sim 0.8$~Myr,
    respectively.}
  \label{fig:hbub}
\end{figure}

The high-energy wind from the AGN begins heating and pushing the
surrounding medium as soon as the AGN switches on. In the spherically
symmetric case, the hot bubble expands and a shell forms around
it. The shell initially expands with a velocity $v_{\rm out} \sim
900$~km/s, consistent with the analytical prediction $v_{\rm e} \simeq
925$~km/s (eq. \ref{eq:ve}; see also Figure \ref{fig:hbub}). Very
quickly, however, the shell cools and starts fragmenting into
filaments and clumps. This effect has already been described in
\citet{Nayakshin2012MNRASb}. A side effect of fragmentation is that
the radial expansion slows down to $\sim 700$~km/s. This occurs partly
due to leaking of wind energy through gaps of lower density and partly
due to the outflowing shell encountering ambient gas with
progressively larger inward velocities. Gas rotation changes some
details of the clump positions, but the overall dynamics of the
outflow are hardly affected.

\begin{figure*}
  \centering
    \includegraphics[width=0.32 \textwidth]{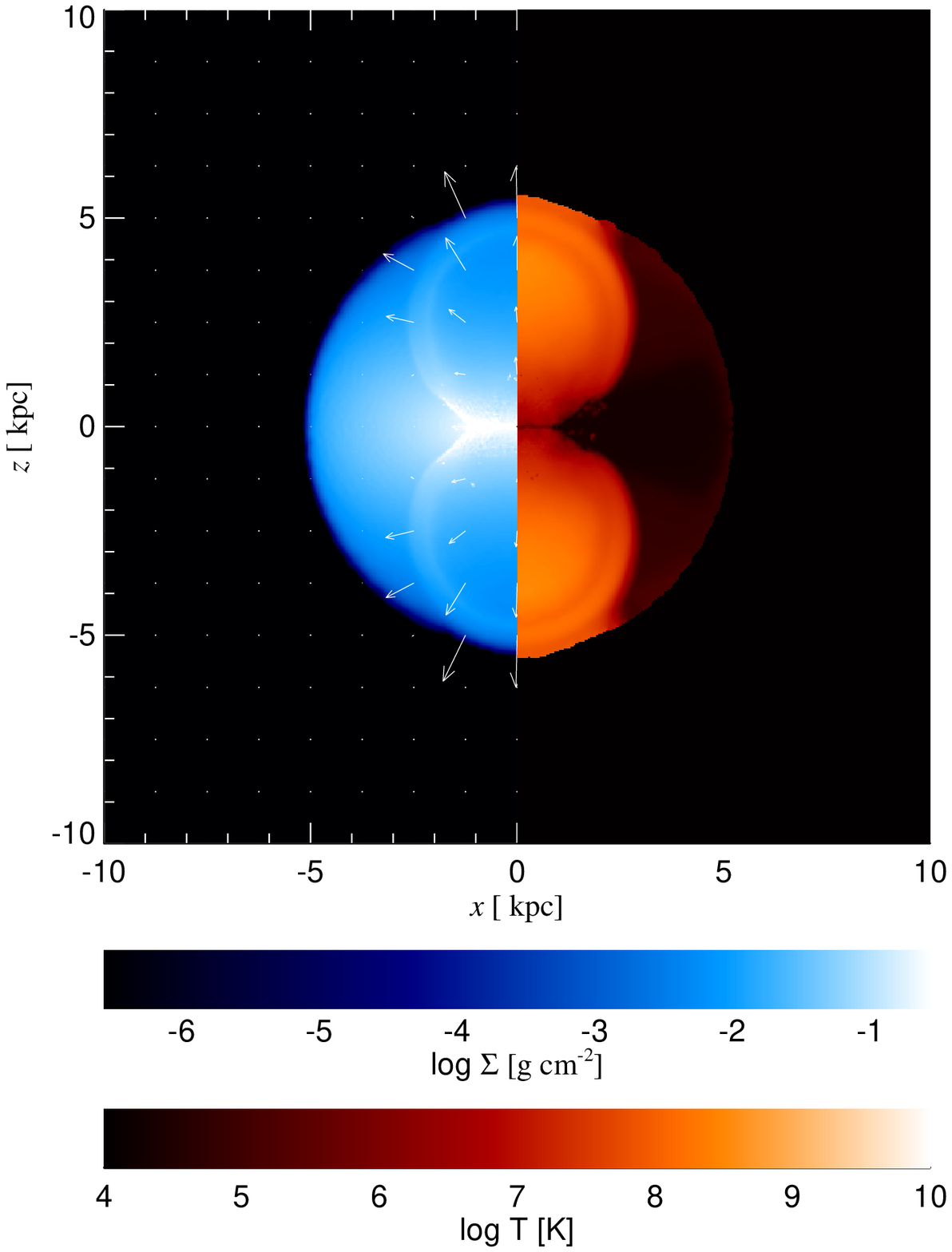}
    \includegraphics[width=0.32 \textwidth]{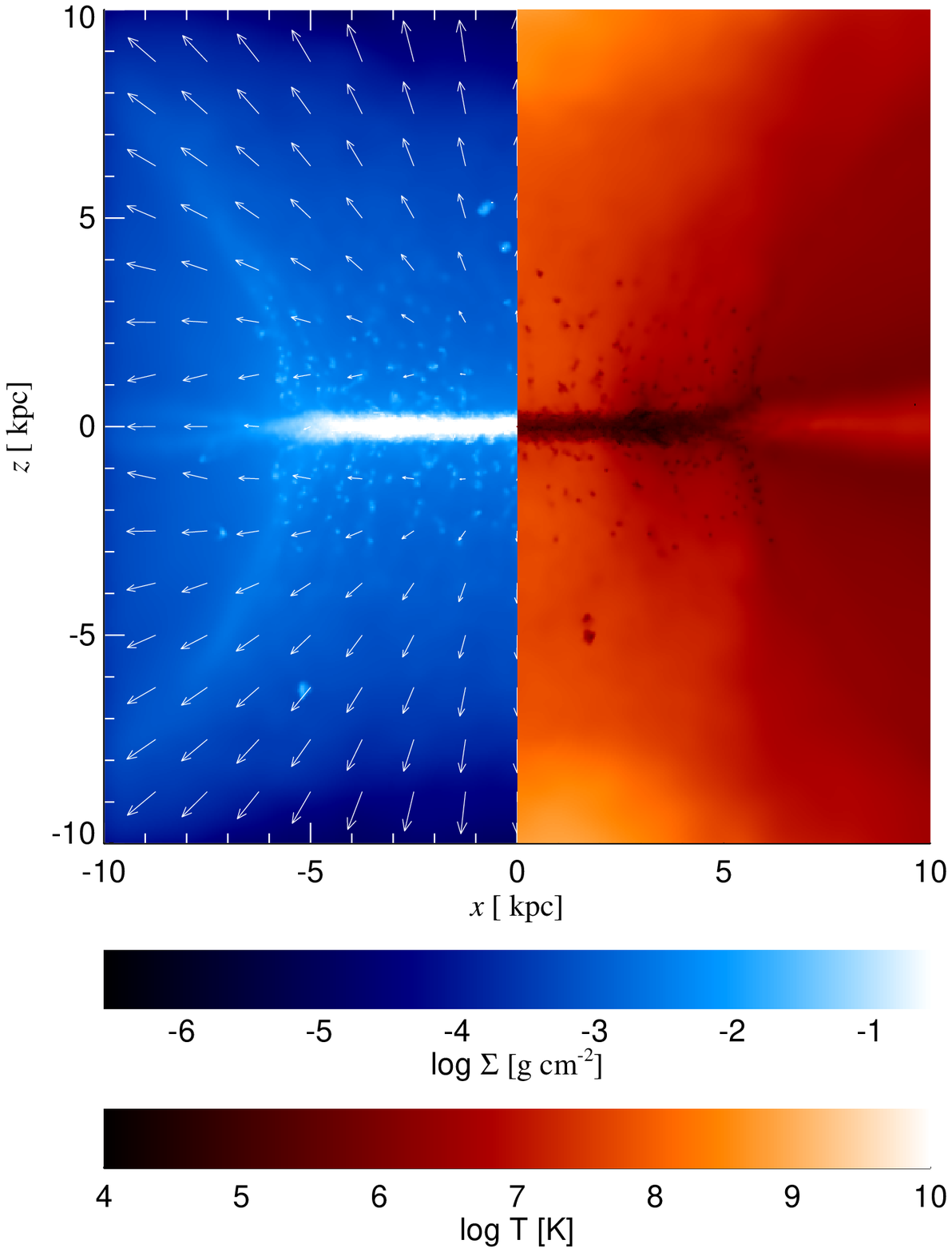}
    \includegraphics[width=0.32 \textwidth]{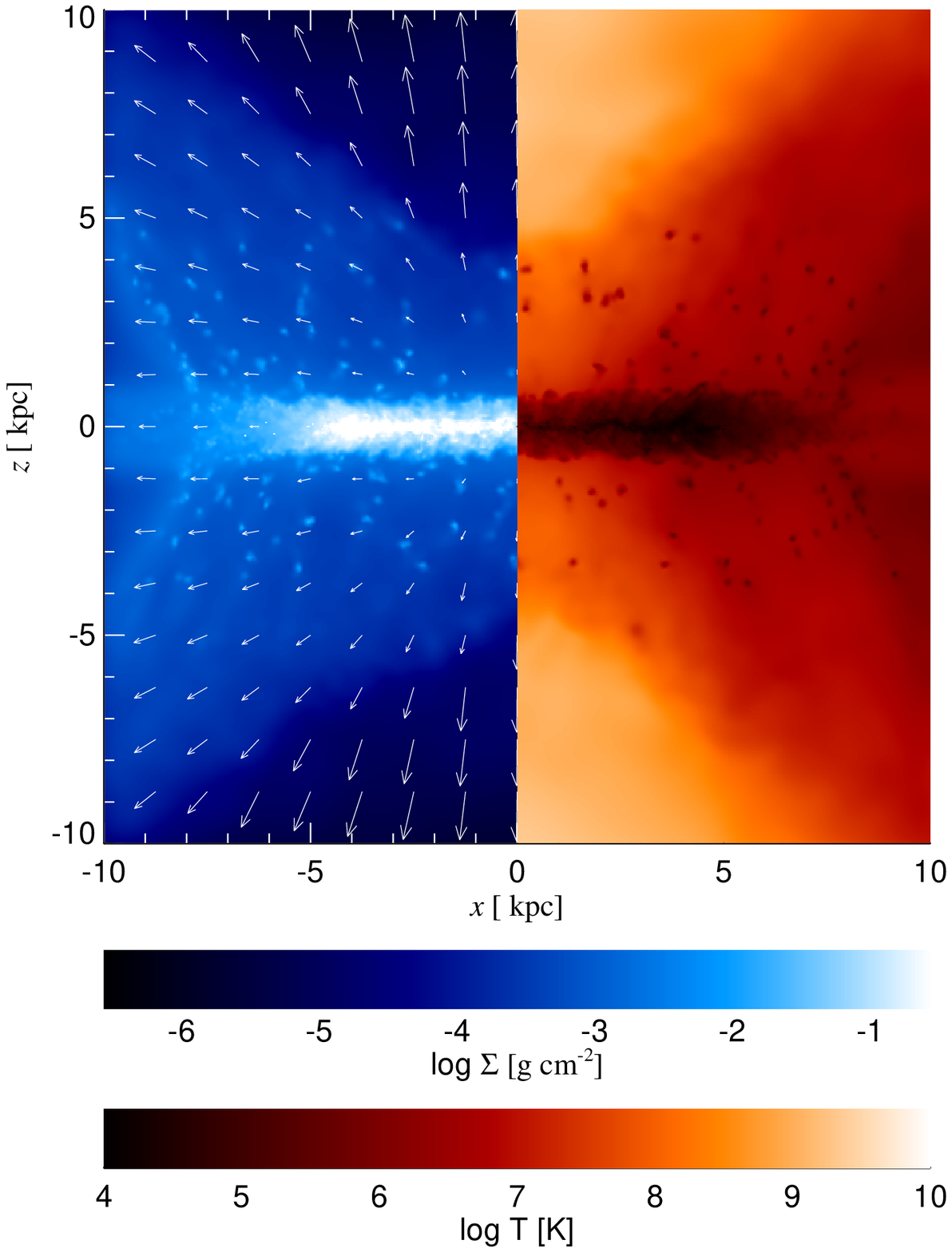}
  \caption{Evolution of an elliptical gas distribution, Z2R0. {\bf
      Left panel:} at $t = 1.3$~Myr, two bubbles have expanded to the
    edge of the initial gas distribution, while in the midplane, gas
    is barely pushed outward at all. {\bf Middle panel:} at $t =
    6.7$~Myr, the midplane gas is compressed in a dense disc, while
    some outflowing gas condenses into denser clumps. {\bf Right
      panel:} at $t = 10$~Myr, the hot bubbles have risen $\sim 3$~kpc
    from the midplane, allowing disc gas to evaporate and create an
    atmosphere of intermediate density.}
  \label{fig:Z2evol}
\end{figure*}

In non-spherical models, the outflow quickly develops a bipolar
morphology (Figure \ref{fig:Z2evol}, left panel). By $1.3$~Myr, the
bubble has already reached the edge of the initial gas distribution
($h_{\rm bubble} \simeq 5$~kpc; see also Figure \ref{fig:hbub}), while
in the equatorial direction, gas is only pushed out to $\sim
0.7$~kpc. The mean velocities in polar and equatorial directions are,
therefore, $\sim 3800$~km/s and $\sim 540$~km/s, respectively. The
analytical calculation, based on initial density contrast, gives
a much smaller difference between the two: $v_{\rm polar} \simeq
1500$~km/s and $v_{\rm equat} \simeq 750$~km/s. This result suggests
that there is significant re-direction of shocked wind energy from the
equator toward the direction of least resistance, i.e. toward the
poles (also see Section \ref{sec:gasdynamics}, below).

Later, the bubble expands laterally as well as vertically, further
compressing the gas in the midplane. Several cold dense clumps form
and move outward, embedded in the hot gas (see Figure
\ref{fig:Z2evol}, middle panel; also see Section
\ref{sec:induced_sf}), but most of the dense gas stays in the
equatorial plane. By 10 Myr (Figure \ref{fig:Z2evol}, right panel),
the cold gas is squeezed into a thin disc with vertical extent $\Delta
z \simeq 1$~kpc. The disc is pushed outwards at a mean velocity of
$\sim 400$ km/s. Its density reaches $10^7$~cm$^{-3}$ and sink
particles start to form. By the end of the simulations, $\sim 2.5\%$
of the SPH particles are converted into sink particles.

One interesting feature developing at late times is an outflowing
``atmosphere'' on either side of the disc. This region of intermediate
density ($n = 0.1-10$~cm$^{-3}$) extends above the disc plane to a
height of $\sim 3-4$~kpc in the central regions, flaring to reach more
than $8$~kpc at cylindrical radii of $10$~kpc.  Gas in this region
rises with vertical velocity $v_{\rm z,atm} \simeq 600-700$~km/s and
has a similar velocity component in the XY plane. Such an outflow
might be erroneously interpreted as a supernova-driven wind coming
from the disc, especially if observed coming from a face-on galaxy
where the planar velocity component is difficult to determine.

\subsection{Gas dynamics} \label{sec:gasdynamics}

\begin{figure}
  \centering
    \includegraphics[width=0.45 \textwidth]{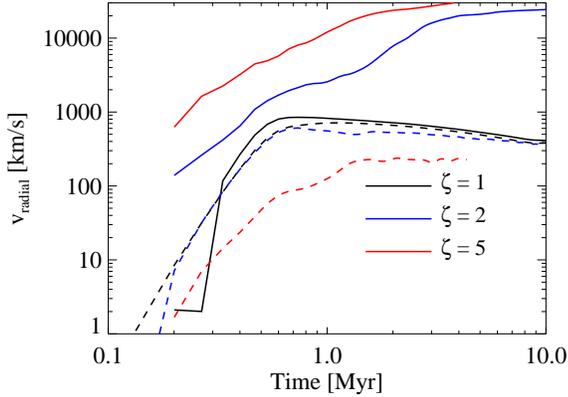}
  \caption{Mean radial velocity of gas in the polar
    ($\left|z\right|/\sqrt{x^2 + y^2} > 4$, solid lines) and
    equatorial ($\left|z\right|/\sqrt{x^2 + y^2} < 0.25$, dashed
    lines) directions for the non-rotating models. Black lines
    represent initially spherically symmetric model Z0R0, blue lines
    represent Z2R0 and red lines Z5R0. As expected, the spherically
    symmetric model has gas expanding with similar velocities of $\sim
    700-800$~km/s in both directions. Outflows in elliptical gas
    distributions, however, are significantly asymmetric: the
    horizontal velocity is similar to or lower than in Z0, while the
    vertical velocities reach several thousand km/s even before the
    outflow bubble escapes the initial gas shell.}
  \label{fig:vradgas}
\end{figure}

In Figure \ref{fig:vradgas}, we plot the mean radial velocity of
outflowing gas as function of time. Solid lines show vertical
expansion velocity (i.e. velocity of gas with
$\left|z\right|/\sqrt{x^2 + y^2} > 4$), while dashed lines show the
radial velocity in the horizontal direction ($\left|z\right|/\sqrt{x^2
  + y^2} < 0.25$). We only plot the velocities of the non-rotating
models; rotating models show very similar behaviour.

As expected, the two velocities in the spherically symmetric case are
approximately the same. The velocity rises to a maximum of
$700-800$~km/s with a broad peak around $t = 1$~Myr, before slowly
decreasing; by $10$~Myr, velocity has dropped to $400-500$~km/s. The
decrease happens for three reasons. First of all, some of the wind
energy leaks out through gaps in the bubble, so the force driving the
outflow decreases. Secondly, the ambient gas shell accelerates inwards
before joining the outflow, so the mass in the outflowing shell
increases slightly faster than linearly (as assumed in the analytical
calculation). The third reason is numerical, as some of the wind
energy is lost due to overcooling of the highest temperature gas.

\begin{figure}
  \centering
    \includegraphics[width=0.45 \textwidth]{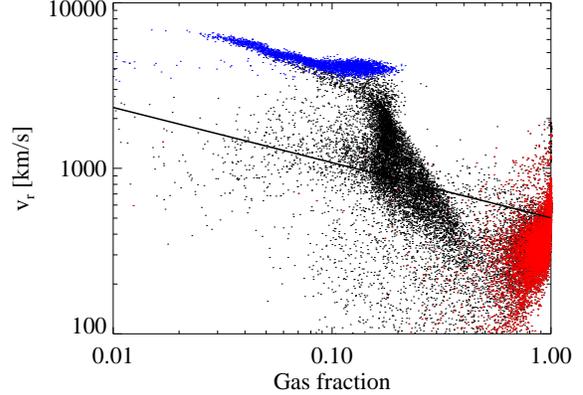}
  \caption{Radial velocities of gas as a function of the effective gas
    fraction ($\rho_{\rm gas}/(\rho_{\rm gas} + \rho_{\rm pot})$) at
    $t = 1.5$~Myr in model Z2R0. Only $5\%$ of all gas particles is
    plotted for clarity. Particles close to the Z axis
    ($\left|z\right|/\sqrt{x^2 + y^2} > 4$) are marked in blue (all
    such particles are plotted, since there are relatively few of
    them) and those close to the midplane ($\left|z\right|/\sqrt{x^2 +
      y^2} < 0.25$) in red. The thick line corresponds to the
    analytical prediction (eq. \ref{eq:ve}). Dense gas tends to be
    slower than the analytical calculation predicts, while diffuse gas
    tends to be faster.}
  \label{fig:veldiag}
\end{figure}

In the non-spherical models, the difference in velocities in the two
directions is significant from the very beginning. In the model Z2R0,
the equatorial plane velocity ($v_{\rm r,hor} \sim 600$~km/s) is
slightly lower than for the spherically symmetric model and the
analytical prediction ($v \sim 750$~km/s). In the vertical direction,
the velocity rapidly increases to $\sim 2500$~km/s (higher than the
analytical prediction of $1540$~km/s). The velocity then stays
approximately constant until the bubble breaks out of the initial
shell. The situation is qualitatively similar in the Z5 model, except
the difference between the analytical prediction and numerical result is
even more pronounced ($v_{\rm r,hor} \simeq 250$~km/s and $v_{\rm
  r,vert} \simeq 10^4$~km/s).

Part of the reason for the difference between analytically calculated
outflow velocities and numerical results is the evolving density
contrast between gas in the polar and equatorial directions. The
initial differences gradually grow as gas is evacuated from the
bubbles in the polar direction, while at the same time the
high-pressure bubbles squeeze the gas vertically toward the
midplane. The net result is that the gas fraction of dense gas is
increasing, and its velocity decreasing, while the opposite is true
for the diffuse gas in the polar direction.

\begin{figure*}
  \centering
    \includegraphics[width=0.45 \textwidth]{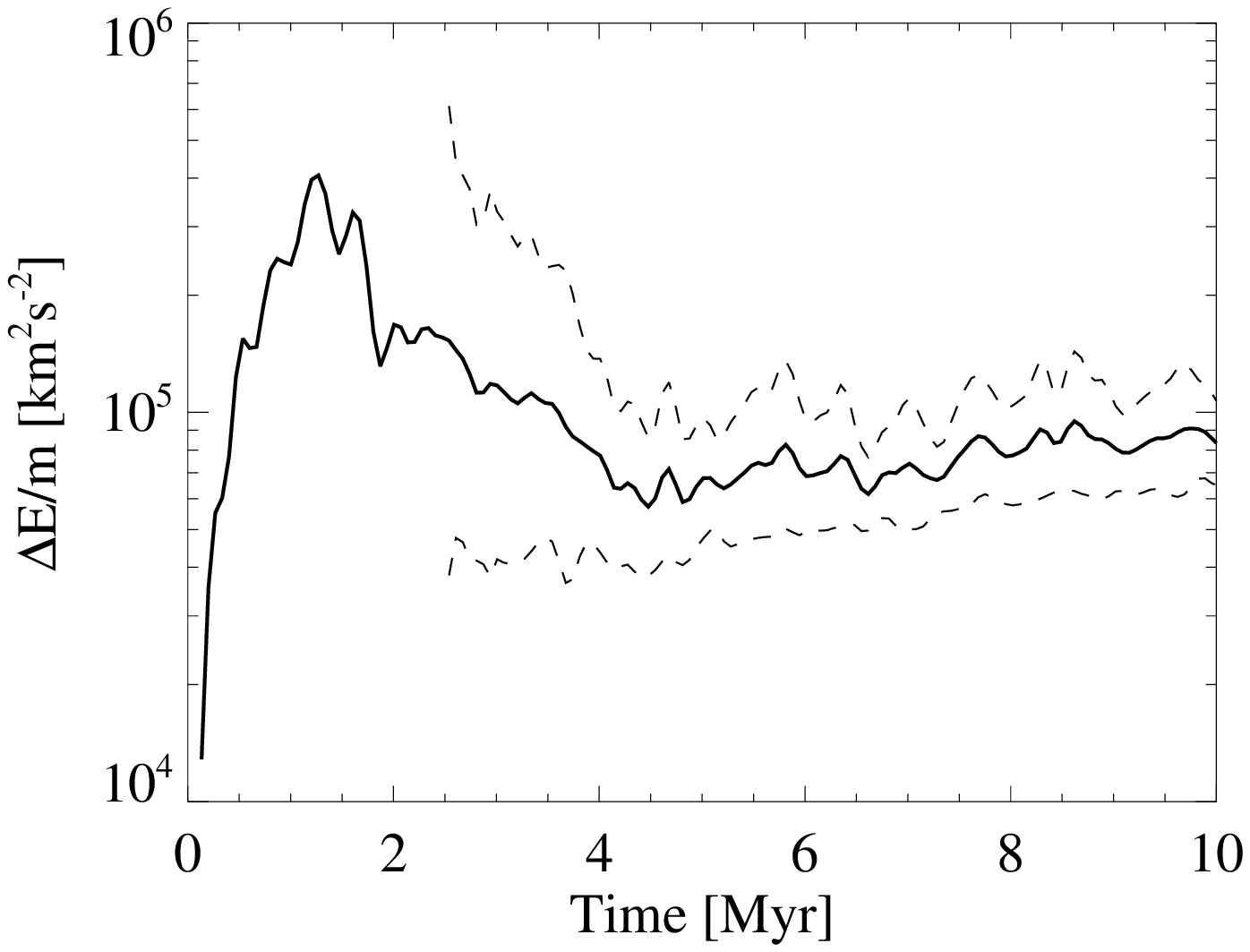}
    \includegraphics[width=0.45 \textwidth]{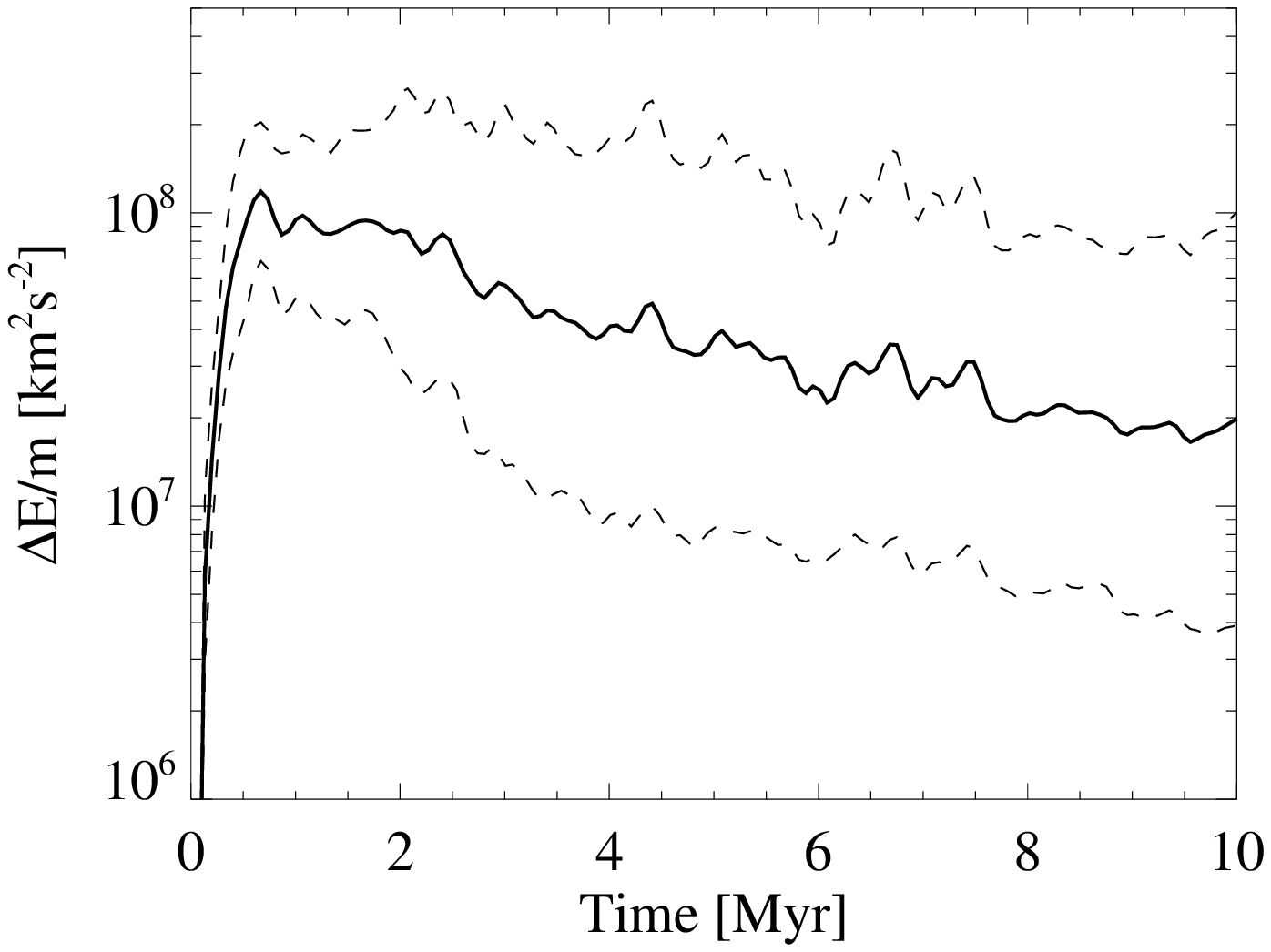}
  \caption{Change in total specific particle energy of dense ($f_{\rm
      g,eff} > 0.1$, left panel) and diffuse ($f_{\rm g,eff} < 0.01$,
    right panel) gas in simulation Z2R0. The solid line is mean energy
    gain, while the dashed lines show $1\sigma$ deviations. Some of
    the dense gas particles have negative change at $t \la 3$~Myr and
    therefore the logarithmic deviation cannot be calculated. Dense
    gas has typical specific energies $\sim 2$ orders of magnitude
    smaller than diffuse gas.}
  \label{fig:histories}
\end{figure*}

However, the leaking of wind energy in the direction of lower
resistance is also an important, if not dominant, contributor to the
velocity difference. To see this, we plot the velocities and effective
gas fractions for a selection of the SPH particles ($5\%$ of the total
number, chosen randomly) in simulation Z2R0 at $t=1.5$ Myr (Figure
\ref{fig:veldiag}). The thick line shows the analytically predicted
velocity (eq. \ref{eq:ve}); we colour particles close to the Z axis
($\left|z\right|/\sqrt{x^2 + y^2} > 4$; we plot all of these
particles, since there are very few of them) in blue and those close
to the midplane ($\left|z\right|/\sqrt{x^2 + y^2} < 0.25$) in red;
these are the same particles as the one used to calculate the mean
vertical and horizontal velocities for Figure \ref{fig:vradgas}. The
diagram shows only a very weak correlation between the analytical
prediction and the actual gas velocities. Most gas with $f_{\rm g,eff}
\ga 0.25$ has velocities lower than predicted (the mean velocity of
this gas is half of the predicted value), while gas with $f_{\rm
  g,eff} \la 0.2$ tends to move faster than predicted (the mean
velocity is $\sim 10\%$ higher than given by eq. \ref{eq:ve}, with
some gas reaching velocities more than three times higher than
predicted). Some gas particles (blue points) move in the vertical
direction with velocities of several thousand km/s; at later times,
they accelerate to a significant fraction of the wind velocity $v_{\rm
  w} = 0.1c$. There are some outliers - gas particles with very high
$f_{\rm g,eff}$, which nevertheless have velocities higher than
predicted analytically. These particles are the clumps seen in Figure
\ref{fig:Z2evol} embedded in the diffuse outflow. They were
accelerated to higher velocities very early in the simulation, when
energy leaking in the vertical direction was less significant, and now
continue to coast with those higher velocities through the lower
density surrounding outflow. Overall, this diagram shows that diffuse
gas carries away a significantly larger fraction of the shocked wind
energy than would be predicted based on the analytical calculation
applicable to the spherically symmetric case. We now show that this
energy leaking results in cold dense gas being pushed mainly by the
momentum of the AGN wind.

\subsection{Energy partition in the flow}

In this section, we show that the distribution of SPH particle
energies reveals that low-density particles carry away most of the
energy, while dense gas is only pushed away by the momentum of the AGN
wind.

First of all, in Figure \ref{fig:histories}, we plot the change in the
SPH particle specific energy for dense ($f_{\rm g,eff} > 0.1$, left
panel) and diffuse ($f_{\rm g,eff} < 0.01$, right panel) gas in
simulation Z2R0. The particle specific energy is defined as the sum of
internal, kinetic and gravitational potential energy:
\begin{equation}\label{eq:specen}
e_{\rm tot} \equiv \frac{E_{\rm tot}}{m_{\rm SPH}} = \frac{3 k_{\rm B}
  T}{2\mu m_{\rm p}} + \frac{v_{\rm r}^2}{2} + 2\sigma^2 {\rm
  ln}\frac{r}{20 {\rm kpc}}.
\end{equation}
Here, the $20$~kpc scaling is chosen as an outer edge of the
background gravitational potential. Self-gravity of the gas is not
included in the definition above, but its contribution is small since
the potential is dominated by the dark matter potential. The quantity
plotted in the graphs is $\Delta e = e_{\rm tot}(t) - e_{\rm tot}(0)$
and is expressed in units of km$^2$s$^{-2}$. Thick solid lines show
the mean energy, while the dashed lines represent $1\sigma$
deviations.

We see that the particle energies stay approximately constant from
$\sim 4$~Myr onward, and that the diffuse gas has $\sim 100$ times
higher specific energy than dense gas.  Moreover, the typical energy
input into dense particles is $\sim (100-300 {\rm km s}^{-1})^2$, of
the same order as expected for a momentum-driven flow ($\Delta v \sim
\sigma$). The energy input into diffuse particles is $\sim (1000-3000
      {\rm km s}^{-1})^2$, slightly higher than expected for a
      spherically symmetric energy-driven flow. This energy
      discrepancy lends support to the claim that low-density
      particles are accelerated preferentially.

\begin{figure}
  \centering
    \includegraphics[width=0.45 \textwidth]{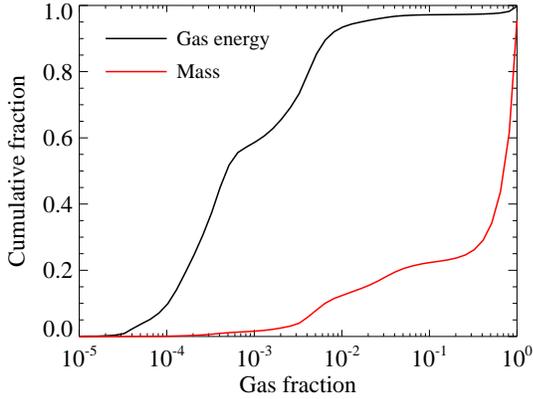}
  \caption{Cumulative distribution of particle energy (black) and mass
    (red) as a function of effective gas fraction ($\rho_{\rm
      gas}/(\rho_{\rm gas} + \rho_{\rm pot})$) at $t = 10$~Myr in
    model Z2R0. The curves are scaled to the total energy and mass,
    respectively. The $5\%$ least dense particles carry $\sim 50\%$ of
    total energy, while the $50\%$ densest particles carry only $<5\%$
    of the energy.}
  \label{fig:en_cumul}
\end{figure}

In Figure \ref{fig:en_cumul}, we plot the cumulative distribution of
the total gas mass (red line) and specific energy gain ($e_{\rm tot} -
e_{\rm tot}(0)$, black line) as function of effective gas fraction at
$t = 10$~Myr for the model Z2R0. Most of the gas mass is contained in
particles with $f_{\rm g,eff} > 0.1$. On the other hand, most of the
energy is carried away by particles with very low densities ($f_{\rm
  g,eff} \la 10^{-3}$). Once again, this shows that energy is removed
by the low density gas, while dense gas is exposed predominantly to
the momentum of the AGN outflow.

\begin{figure}
  \centering
    \includegraphics[width=0.45 \textwidth]{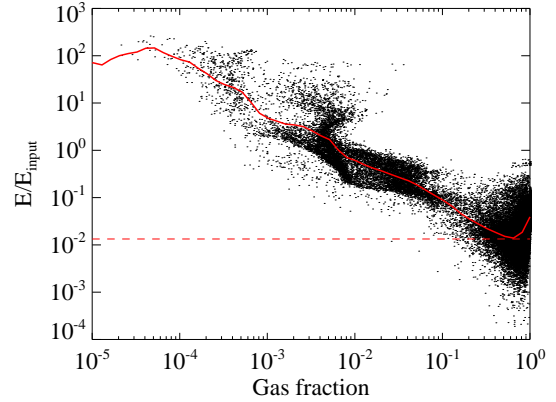}
  \caption{Ratio between SPH particle specific energy gain, $\Delta e$
    (eq. \ref{eq:specen}), and the mean specific energy input from the
    AGN, as a function of the effective gas fraction ($\rho_{\rm
      gas}/(\rho_{\rm gas} + \rho_{\rm pot})$) at $t = 1.3$~Myr in
    model Z2R0. Only $5\%$ of all SPH particles are plotted. The thick
    red line shows the mean logarithmic value at each gas
    fraction. The thin dashed line is the analytical prediction of
    input energy retained in a momentum-driven flow ($f_{\rm mom} = 2
    \sigma / (\eta c) = 0.013$). Dense gas ($f_{\rm g,eff} \ga 0.1$)
    has mean energy similar to the momentum-driven outflow
    solution. The slight increase for the densest gas occurs because
    particle energy is overestimated due to not accounting for
    self-gravity of the gas, and also because some of this gas is in
    more rapidly moving clumps.}
  \label{fig:en_ratio}
\end{figure}

In Figure \ref{fig:en_ratio} we plot the ratio of SPH particle energy
gain to the AGN wind energy input per particle as function of
effective gas fraction for simulation Z2R0 at $t = 10$~Myr. The AGN
wind energy input is simply $E_{\rm in} = \eta L_{\rm Edd} t /(2
N_{\rm SPH})$. The thick red line is the mean of the logarithm of this
energy ratio (a small number, $\sim 500, of$ particles with negative
energy gain are not included when calculating the mean), while the
thin dashed line shows the fraction of input energy retained in a
purely momentum-driven flow ($f_{\rm mom} = 2 \sigma / (\eta c) =
0.013$).

Once again, we see that the low-density gas has much more energy than
its share of the input. More revealing, the dense gas ($f_{\rm g,eff}
\ga 0.1$) has a typical specific energy retention fraction of
$0.015-0.04$, close to the momentum-driven outflow prediction. The
fact that the typical energy ratio in dense gas varies by only a
factor of $\la 3$ lends support to the conclusion that dense gas is
driven away by AGN wind momentum rather than its energy. The densest
gas with slightly higher energies is generally gravitationally bound,
and therefore its energy is slightly overestimated in our analysis.

\section{Discussion}\label{sec:discuss}

\subsection{Summary of main results}

Our simulations of AGN outflows in non-spherical geometries reveal
three important effects. First of all, the outflow expands in two
elongated bubbles in the direction of least resistance, compressing
the denser midplane gas into a disc. Secondly, the AGN fast wind
outflow predominantly works on the least dense gas, increasing its
velocity significantly above that calculated analytically assuming
that outflows of ambient gas along different directions are
decoupled. Conversely, the dense gas is mostly affected by the ram
pressure of the outflow (again contrary to the simplest analytical
spherically symmetric model with no energy leakage). Finally, at late
times the outflowing bubble rises up from the disc, allowing the disc
to evaporate and create an outflowing atmosphere of intermediate
density between the disc and the bubble.

These results have implications for the establishment of $M-\sigma$
relation via self-regulation of SMBH feeding. In addition, gas
compression implies that star formation can be triggered by the
outflow. Finally, the morphology of the outflows suggests that certain
observed galaxy properties - namely the presence of bubbles or winds
rising from discs - can be driven by wide-angle AGN outflows rather
than other processes. We discuss each of these implications in turn.

\subsection{$M-\sigma$ relation}

The $M-\sigma$ relation derived analytically assumes that the AGN
outflow is momentum-driven, at least out to distances several times
larger than the sphere of influence of the SMBH \citep[also see
  Introduction]{King2010MNRASa}. More detailed calculations showing
that cooling of the wind shock is very inefficient call for
significant modifications to this spherically symmetric
derivation. Our simulation results show that the energy-driven outflow
produces ``momentum-like'' feedback upon dense gas. Dense gas is
pushed away with velocities not much greater than those expected from
analytical calculations of momentum-driven winds. Note that since
radiative cooling in the ambient gas is efficient in conditions
considered here, a large fraction of the gas is dense and hence moves
with low velocity; at $t=1.3$~Myr in simulation Z2R0, half of the
outflowing gas has velocities $v < 470$~km/s.

We also presented an analytical toy model that allows for the bubble
energy escape through low density regions of the shell. This model
shows that the critical SMBH mass for such ``leaky shells'' may in
fact be close to the result obtained by \cite{King2003ApJ} for
spherically symmetric momentum-driven shells. Further numerical work
with initial conditions less idealised than used here is needed to
test this conclusion numerically.

\subsection{Induced star formation} \label{sec:induced_sf}

The bipolar outflow is significantly over-pressurized with respect to
its surroundings, and therefore expands laterally. This causes the
dense gas to be compacted further in the midplane. In simulation Z2R0,
diffuse gas filling the bubble has pressure $P_{\rm hot}/k_{\rm B}
\simeq 10^7 - 10^8$~K~cm$^{-3}$, significantly higher than the typical
ISM pressure $\sim 3 \times 10^5$~K~cm$^{-3}$ \citep{Wolfire2003ApJ}.
This has a strong effect on the gas in the disc. In our simulations,
the disc and its atmosphere have even higher pressures $\ga
10^9$~K~cm$^{-3}$, but this is most likely overestimated by several
orders of magnitude due to the adopted temperature floor. Such
environments are likely to form stars much more rapidly than typical
star-forming regions \citep{Zubovas2013MNRASb, Krumholz2009ApJ}. Our
simulations do not include the relevant physics of star formation
feedback, and hence may over-estimate the corresponding star formation
rates. Therefore we do not provide quantitative predictions of the
effect the outflow has upon the SFR in the galactic disc. We note that
\citet{Gaibler2012MNRAS} found a similar over-pressurizing effect
caused by a jet-inflated bubble.

Dense gas also appears in the form of clumps throughout the outflow
(see Figure \ref{fig:Z2evol}, middle and right
panels). Morphologically, these clumps are similar to the
high-velocity clouds commonly seen in galaxy formation simulations,
which have been recently identified as numerical artifacts of standard
SPH formulations \citep{Hobbs2013MNRAS}. However, we are confident
that the presence of clumps in our simulations is a robust conclusion,
even if the longevity of individual clumps is overestimated. We think
so because of an earlier paper \citep{Nayakshin2012MNRASb}, where we
showed that these clumps form due to a combination of efficient gas
cooling and gravitational instability in the outflow. In addition,
\citet{Zubovas2014arXiv} showed analytically that the formation of
cold gas is a very rapid process, and most of the outflowing gas
should become molecular. Therefore cold dense gas is expected to exist
in the outflow.

\subsection{Outflow bubbles -- fake jets?}

Evidence of bipolar outflows from the centres of galaxies is
well-known both in the Milky Way \citep[the {\em Fermi}
  bubbles;][]{Su2010ApJ} and in active galaxies
\citep[e.g.,][]{Storchi2007ApJ,Veilleux2001AJ}. Quite often, the
collimation of the bubbles is taken as evidence that the bubbles are
produced by jets \citep{Nesvadba2008A&A,Guo2012ApJ}. Our results show
that in fact, a spherically symmetric outflow can be efficiently
collimated by the uneven density distribution in the host galaxy and
produce elongated bubbles. These bubbles should be intrinsically
bright gamma-ray sources due to multiple shock fronts within them
\citep{Zubovas2012MNRASa, King2010MNRASb} and so could present
observational features similar to those of bubbles inflated by
jets. This suggests that the mere presence of an outflow bubble does
not necessarily indicate past jet activity of the AGN, and that
estimates of jet power based on the mechanical power of the bubbles
should be taken as upper limits only.

Bipolar bubbles inflated by initially spherical outflows are also a
common effect of stellar feedback, mostly supernova explosions
\citep{MacLow1999ApJ,Tenorio1998MNRAS,Springel2003MNRASb}, especially
in low-mass galaxies \citep{Dubois2008A&A}. The AGN-wind-driven
outflow bubbles are a natural extension of these models into a higher
mass and higher energy regime.

\subsection{Pseudo-disc outflows and galactic fountains}

The disc ``atmosphere'', seen in the non-spherical simulations,
contains gas moving with velocities of several hundred km/s in the
vertical direction (i.e. directly away from the plane). Such
velocities are sometimes large enough for the gas to escape the galaxy
in a pseudo-disc wind (we call it ``pseudo'' because it is not
triggered by any process happening in the disc). Other parcels of gas
would later fall back on the disc further from the centre,
contributing to the galactic fountain.

Disc winds with large velocities ($\ga 500$~km/s) have been observed
in many galaxies \citep{Shopbell1998ApJ,Coil2011ApJ}. These velocities
are difficult to achieve by supernova driving alone
\citep{Coil2011ApJ}, and when they are, only a small fraction of gas
can be accelerated to such large velocities
\citep{Strickland2000MNRAS}. Our results suggest that some of these
outflows can be caused, fully or in part, by AGN activity. These
galaxies may have experienced episodes of AGN activity in the past,
and one footprint of this activity is a vertically rebounding disc
gas, which launches rapid outflows independently of star
formation. Starbursts in such discs may also have been triggered by
previous compression due to the AGN outflow.

\section{Conclusion}\label{sec:concl}

Our numerical experiments showed that energy-conserving
spherically-symmetric outflows from SMBHs may create highly aspherical
bubbles if the ambient gas in the host galaxy is not spherically
distributed. This may lead to a whole host of theoretical and
observational implications for SMBH-host galaxy connections.

Firstly, SMBH driving energy-conserving outflows may self-regulate
their growth to the momentum-conserving $M_\sigma$ value found by
\cite{King2003ApJ}. This implies that SMBH outflows may actually not
lose much energy to radiation, as assumed in the momentum-conserving
picture, and be therefore pumping {\em all} of their energy into the
surrounding ambient gas. As we argued here, however, most of this
energy leaks out from the bulge via low density channels and is
therefore deposited outside of the bulge, in the halo of the host
galaxy or even beyond. This leads to dense gas only being exposed to
the ram pressure of the outflow, and thus the critical SMBH mass
required to push it away and halt further accretion is very similar to
$M_\sigma$.

Secondly, the observational appearance of SMBH feedback may be
deceiving. This point is very important: the interpretation of
observed feedback processes in the host galaxies may be incorrect in
some cases. For example, wide angle outflow studied here forms
biconical structures which may look very much like galaxy-disc
outflows driven by starbursts in the discs of the host
galaxies. Another potential mis-interpretation of observations could
be the buoyant bubbles usually presumed to be inflated by AGN jet
activity. The bubbles we obtained here are similarly energetic and are
also filled with very high temperature gas where electrons could be
potentially accelerated into non-thermal distributions in shocks
\citep[as suggested for the Fermi Bubbles by][]{Zubovas2012MNRASa}. If
a jet, however weak, was also present in addition to the wide angle
outflows studied here, then it could be a simple matter to attribute
the bubble's mechanical power to what is easier to discern in the
observations -- the jet.

\section*{Acknowledgments}

We thank the anonymous referee for extensive comments, which helped
significantly improve the clarity of the paper. KZ acknowledges the UK
STFC for support successively in the form of a PhD studentship and a
postdoctoral research position, both at the University of
Leicester. This research is partially supported by the Research
Council Lithuania grant no. MIP-062/2013.

Numerical simulations presented in this work were carried out on two
computing clusters. Some computations were performed on resources at
the High Performance Computing Center HPC Sauletekis in Vilnius
University Faculty of Physics. This work also used the DiRAC
Complexity system, operated by the University of Leicester, which
forms part of the STFC DiRAC HPC Facility (www.dirac.ac.uk). This
equipment is funded by a BIS National E-Infrastructure capital grant
ST/K000373/1 and DiRAC Operations grant ST/K0003259/1. DiRAC is part
of the UK National E-Infrastructure.


\begin{thebibliography}{86}
\expandafter\ifx\csname natexlab\endcsname\relax\def\natexlab#1{#1}\fi

\bibitem[{Aller} \& {Richstone}(2007)]{Aller2007ApJ}
{Aller} M.~C., {Richstone} D.~O., 2007, \apj, 665, 120

\bibitem[{Bailin} \& {Steinmetz}(2005)]{Bailin2005ApJ}
{Bailin} J., {Steinmetz} M., 2005, \apj, 627, 647

\bibitem[{Booth} \& {Schaye}(2009)]{Booth2009MNRAS}
{Booth} C.~M., {Schaye} J., 2009, \mnras, 398, 53

\bibitem[{Bourne} \& {Nayakshin}(2013)]{Bourne2013arXiv}
{Bourne} M.~A., {Nayakshin} S., 2013, ArXiv e-prints

\bibitem[{Bower} et~al.(2006){Bower}, {Benson}, {Malbon}
  et~al.]{Bower2006MNRAS}
{Bower} R.~G., {Benson} A.~J., {Malbon} R., et~al., 2006, \mnras, 370, 645

\bibitem[{Churazov} et~al.(2005){Churazov}, {Sazonov}, {Sunyaev}, {Forman},
  {Jones} \& {B{\"o}hringer}]{ChurazovEtal05}
{Churazov} E., {Sazonov} S., {Sunyaev} R., {Forman} W., {Jones} C.,
  {B{\"o}hringer} H., 2005, \mnras, 363, L91

\bibitem[{Ciotti} \& {Ostriker}(1997)]{Ciotti1997ApJ}
{Ciotti} L., {Ostriker} J.~P., 1997, \apjl, 487, L105+

\bibitem[{Ciotti} \& {Ostriker}(2007)]{Ciotti2007ApJ}
{Ciotti} L., {Ostriker} J.~P., 2007, \apj, 665, 1038

\bibitem[{Coil} et~al.(2011){Coil}, {Weiner}, {Holz}, {Cooper}, {Yan} \&
  {Aird}]{Coil2011ApJ}
{Coil} A.~L., {Weiner} B.~J., {Holz} D.~E., {Cooper} M.~C., {Yan} R., {Aird}
  J., 2011, \apj, 743, 46

\bibitem[{Croton} et~al.(2006){Croton}, {Springel}, {White}
  et~al.]{Croton2006MNRAS}
{Croton} D.~J., {Springel} V., {White} S.~D.~M., et~al., 2006, \mnras, 365, 11

\bibitem[{Daddi} et~al.(2010){Daddi}, {Bournaud}, {Walter}
  et~al.]{Daddi2010ApJ}
{Daddi} E., {Bournaud} F., {Walter} F., et~al., 2010, \apj, 713, 686

\bibitem[{Debuhr} et~al.(2011){Debuhr}, {Quataert} \& {Ma}]{DebuhrEtal11}
{Debuhr} J., {Quataert} E., {Ma} C.-P., 2011, \mnras, 412, 1341

\bibitem[{Debuhr} et~al.(2010){Debuhr}, {Quataert}, {Ma} \&
  {Hopkins}]{Debuhr2010MNRAS}
{Debuhr} J., {Quataert} E., {Ma} C.-P., {Hopkins} P., 2010, \mnras, 406, L55

\bibitem[{Di Matteo} et~al.(2008){Di Matteo}, {Colberg}, {Springel},
  {Hernquist} \& {Sijacki}]{DiMatteo2008ApJ}
{Di Matteo} T., {Colberg} J., {Springel} V., {Hernquist} L., {Sijacki} D.,
  2008, \apj, 676, 33

\bibitem[{Di Matteo} et~al.(2005){Di Matteo}, {Springel} \&
  {Hernquist}]{DiMatteo2005Natur}
{Di Matteo} T., {Springel} V., {Hernquist} L., 2005, \nat, 433, 604

\bibitem[{Diemand} \& {Moore}(2011)]{Diemand2011ASL}
{Diemand} J., {Moore} B., 2011, Advanced Science Letters, 4, 297

\bibitem[{Dubois} et~al.(2012{\natexlab{a}}){Dubois}, {Devriendt}, {Slyz} \&
  {Teyssier}]{Dubois2012MNRAS}
{Dubois} Y., {Devriendt} J., {Slyz} A., {Teyssier} R., 2012{\natexlab{a}},
  \mnras, 420, 2662

\bibitem[{Dubois} et~al.(2012{\natexlab{b}}){Dubois}, {Pichon}, {Haehnelt}
  et~al.]{Dubois2012MNRASb}
{Dubois} Y., {Pichon} C., {Haehnelt} M., et~al., 2012{\natexlab{b}}, \mnras,
  423, 3616

\bibitem[{Dubois} \& {Teyssier}(2008)]{Dubois2008A&A}
{Dubois} Y., {Teyssier} R., 2008, \aap, 477, 79

\bibitem[{Everett} \& {Ballantyne}(2004)]{Everett2004ApJ}
{Everett} J.~E., {Ballantyne} D.~R., 2004, \apjl, 615, L13

\bibitem[{Fabian}(1999)]{Fabian1999MNRAS}
{Fabian} A.~C., 1999, \mnras, 308, L39

\bibitem[{Fabian}(2012)]{Fabian2012ARA&A}
{Fabian} A.~C., 2012, \araa, 50, 455

\bibitem[{Faucher-Gigu{\`e}re} \& {Quataert}(2012)]{Faucher2012MNRASb}
{Faucher-Gigu{\`e}re} C.-A., {Quataert} E., 2012, \mnras, 425, 605

\bibitem[{Feoli} \& {Mancini}(2009)]{Feoli2009ApJ}
{Feoli} A., {Mancini} L., 2009, \apj, 703, 1502

\bibitem[{Ferrarese} \& {Merritt}(2000)]{Ferrarese2000ApJ}
{Ferrarese} L., {Merritt} D., 2000, \apjl, 539, L9

\bibitem[{Feruglio} et~al.(2010){Feruglio}, {Maiolino}, {Piconcelli}
  et~al.]{Feruglio2010A&A}
{Feruglio} C., {Maiolino} R., {Piconcelli} E., et~al., 2010, \aap, 518, L155+

\bibitem[{Gaibler} et~al.(2012){Gaibler}, {Khochfar}, {Krause} \&
  {Silk}]{Gaibler2012MNRAS}
{Gaibler} V., {Khochfar} S., {Krause} M., {Silk} J., 2012, \mnras, 425, 438

\bibitem[{G{\"u}ltekin} et~al.(2009){G{\"u}ltekin}, {Richstone}, {Gebhardt}
  et~al.]{Gultekin2009ApJ}
{G{\"u}ltekin} K., {Richstone} D.~O., {Gebhardt} K., et~al., 2009, \apj, 698,
  198

\bibitem[{Guo} \& {Mathews}(2012)]{Guo2012ApJ}
{Guo} F., {Mathews} W.~G., 2012, \apj, 756, 181

\bibitem[{H{\"a}ring} \& {Rix}(2004)]{Haering2004ApJ}
{H{\"a}ring} N., {Rix} H.-W., 2004, \apjl, 604, L89

\bibitem[{Harper-Clark} \& {Murray}(2009)]{HMurray09}
{Harper-Clark} E., {Murray} N., 2009, \apj, 693, 1696

\bibitem[{Hobbs} et~al.(2013{\natexlab{a}}){Hobbs}, {Read}, {Power} \&
  {Cole}]{HobbsEtal13a}
{Hobbs} A., {Read} J., {Power} C., {Cole} D., 2013{\natexlab{a}}, \mnras, 434,
  1849

\bibitem[{Hobbs} et~al.(2013{\natexlab{b}}){Hobbs}, {Read}, {Power} \&
  {Cole}]{Hobbs2013MNRAS}
{Hobbs} A., {Read} J., {Power} C., {Cole} D., 2013{\natexlab{b}}, \mnras, 434,
  1849

\bibitem[{King}(2003)]{King2003ApJ}
{King} A., 2003, \apjl, 596, L27

\bibitem[{King}(2005)]{King2005ApJ}
{King} A., 2005, \apjl, 635, L121

\bibitem[{King} et~al.(2013){King}, {Miller}, {Raymond} et~al.]{King2013ApJ}
{King} A.~L., {Miller} J.~M., {Raymond} J., et~al., 2013, \apj, 762, 103

\bibitem[{King}(2010{\natexlab{a}})]{King2010MNRASb}
{King} A.~R., 2010{\natexlab{a}}, \mnras, 408, L95

\bibitem[{King}(2010{\natexlab{b}})]{King2010MNRASa}
{King} A.~R., 2010{\natexlab{b}}, \mnras, 402, 1516

\bibitem[{King} \& {Pounds}(2003)]{King2003MNRASb}
{King} A.~R., {Pounds} K.~A., 2003, \mnras, 345, 657

\bibitem[{King} et~al.(2011){King}, {Zubovas} \& {Power}]{King2011MNRAS}
{King} A.~R., {Zubovas} K., {Power} C., 2011, \mnras,  L263+

\bibitem[{Koopmans} et~al.(2009){Koopmans}, {Bolton}, {Treu}
  et~al.]{Koopmans2009ApJ}
{Koopmans} L.~V.~E., {Bolton} A., {Treu} T., et~al., 2009, \apjl, 703, L51

\bibitem[{Krumholz} et~al.(2009){Krumholz}, {McKee} \&
  {Tumlinson}]{Krumholz2009ApJ}
{Krumholz} M.~R., {McKee} C.~F., {Tumlinson} J., 2009, \apj, 699, 850

\bibitem[{Krumholz} \& {Thompson}(2013)]{KrumholzThompson13}
{Krumholz} M.~R., {Thompson} T.~A., 2013, \mnras, 434, 2329

\bibitem[{Mac Low} \& {Ferrara}(1999)]{MacLow1999ApJ}
{Mac Low} M.-M., {Ferrara} A., 1999, \apj, 513, 142

\bibitem[{Magorrian} et~al.(1998){Magorrian}, {Tremaine}, {Richstone}
  et~al.]{Magorrian1998AJ}
{Magorrian} J., {Tremaine} S., {Richstone} D., et~al., 1998, \aj, 115, 2285

\bibitem[{McConnell} et~al.(2011){McConnell}, {Ma}, {Gebhardt}
  et~al.]{McConnell2011Natur}
{McConnell} N.~J., {Ma} C.-P., {Gebhardt} K., et~al., 2011, \nat, 480, 215

\bibitem[{McQuillin} \& {McLaughlin}(2013)]{McQuillin2013MNRAS}
{McQuillin} R.~C., {McLaughlin} D.~E., 2013, \mnras, 434, 1332

\bibitem[{Murray} et~al.(2005){Murray}, {Quataert} \&
  {Thompson}]{Murray2005ApJ}
{Murray} N., {Quataert} E., {Thompson} T.~A., 2005, \apj, 618, 569

\bibitem[{Nayakshin}(2013)]{Nayakshin13b}
{Nayakshin} S., 2013, \mnras

\bibitem[{Nayakshin} et~al.(2009){Nayakshin}, {Cha} \&
  {Hobbs}]{Nayakshin2009MNRASb}
{Nayakshin} S., {Cha} S.-H., {Hobbs} A., 2009, \mnras, 397, 1314

\bibitem[{Nayakshin} \& {Zubovas}(2012)]{Nayakshin2012MNRASb}
{Nayakshin} S., {Zubovas} K., 2012, \mnras, 427, 372

\bibitem[{Nesvadba} et~al.(2008){Nesvadba}, {Lehnert}, {De Breuck}, {Gilbert}
  \& {van Breugel}]{Nesvadba2008A&A}
{Nesvadba} N.~P.~H., {Lehnert} M.~D., {De Breuck} C., {Gilbert} A.~M., {van
  Breugel} W., 2008, \aap, 491, 407

\bibitem[{Page} et~al.(2012){Page}, {Symeonidis}, {Vieira}
  et~al.]{Page2012Natur}
{Page} M.~J., {Symeonidis} M., {Vieira} J.~D., et~al., 2012, \nat, 485, 213

\bibitem[{Pounds} et~al.(2003{\natexlab{a}}){Pounds}, {King}, {Page} \&
  {O'Brien}]{Pounds2003MNRASb}
{Pounds} K.~A., {King} A.~R., {Page} K.~L., {O'Brien} P.~T.,
  2003{\natexlab{a}}, \mnras, 346, 1025

\bibitem[{Pounds} et~al.(2003{\natexlab{b}}){Pounds}, {Reeves}, {King}, {Page},
  {O'Brien} \& {Turner}]{Pounds2003MNRASa}
{Pounds} K.~A., {Reeves} J.~N., {King} A.~R., {Page} K.~L., {O'Brien} P.~T.,
  {Turner} M.~J.~L., 2003{\natexlab{b}}, \mnras, 345, 705

\bibitem[{Pounds} \& {Vaughan}(2011{\natexlab{a}})]{Pounds2011MNRAS}
{Pounds} K.~A., {Vaughan} S., 2011{\natexlab{a}}, \mnras, 413, 1251

\bibitem[{Pounds} \& {Vaughan}(2011{\natexlab{b}})]{PoundsVaughan11}
{Pounds} K.~A., {Vaughan} S., 2011{\natexlab{b}}, \mnras, 415, 2379

\bibitem[{Proga} et~al.(2000){Proga}, {Stone} \& {Kallman}]{Proga2000ApJ}
{Proga} D., {Stone} J.~M., {Kallman} T.~R., 2000, \apj, 543, 686

\bibitem[{Rupke} \& {Veilleux}(2011)]{Rupke2011ApJ}
{Rupke} D.~S.~N., {Veilleux} S., 2011, \apjl, 729, L27+

\bibitem[{Santini} et~al.(2014){Santini}, {Maiolino}, {Magnelli}
  et~al.]{Santini2014A&A}
{Santini} P., {Maiolino} R., {Magnelli} B., et~al., 2014, \aap, 562, A30

\bibitem[{Sazonov} et~al.(2005){Sazonov}, {Ostriker}, {Ciotti} \&
  {Sunyaev}]{Sazonov2005MNRAS}
{Sazonov} S.~Y., {Ostriker} J.~P., {Ciotti} L., {Sunyaev} R.~A., 2005, \mnras,
  358, 168

\bibitem[{Shakura} \& {Sunyaev}(1973)]{Shakura1973A&A}
{Shakura} N.~I., {Sunyaev} R.~A., 1973, \aap, 24, 337

\bibitem[{Shopbell} \& {Bland-Hawthorn}(1998)]{Shopbell1998ApJ}
{Shopbell} P.~L., {Bland-Hawthorn} J., 1998, \apj, 493, 129

\bibitem[{Sijacki} et~al.(2007){Sijacki}, {Springel}, {Di Matteo} \&
  {Hernquist}]{Sijacki2007MNRAS}
{Sijacki} D., {Springel} V., {Di Matteo} T., {Hernquist} L., 2007, \mnras, 380,
  877

\bibitem[{Silk} \& {Nusser}(2010)]{SilkNusser10}
{Silk} J., {Nusser} A., 2010, \apj, 725, 556

\bibitem[{Silk} \& {Rees}(1998)]{Silk1998A&A}
{Silk} J., {Rees} M.~J., 1998, \aap, 331, L1

\bibitem[{Springel}(2005)]{Springel2005MNRAS}
{Springel} V., 2005, \mnras, 364, 1105

\bibitem[{Springel}(2010)]{Springel2010ARA&A}
{Springel} V., 2010, \araa, 48, 391

\bibitem[{Springel} \& {Hernquist}(2003)]{Springel2003MNRASb}
{Springel} V., {Hernquist} L., 2003, \mnras, 339, 312

\bibitem[{Storchi-Bergmann} et~al.(2007){Storchi-Bergmann}, {Dors}, {Riffel}
  et~al.]{Storchi2007ApJ}
{Storchi-Bergmann} T., {Dors} Jr. O.~L., {Riffel} R.~A., et~al., 2007, \apj,
  670, 959

\bibitem[{Strickland} \& {Stevens}(2000)]{Strickland2000MNRAS}
{Strickland} D.~K., {Stevens} I.~R., 2000, \mnras, 314, 511

\bibitem[{Sturm} et~al.(2011){Sturm}, {Gonz{\'a}lez-Alfonso}, {Veilleux}
  et~al.]{Sturm2011ApJ}
{Sturm} E., {Gonz{\'a}lez-Alfonso} E., {Veilleux} S., et~al., 2011, \apjl, 733,
  L16+

\bibitem[{Su} et~al.(2010){Su}, {Slatyer} \& {Finkbeiner}]{Su2010ApJ}
{Su} M., {Slatyer} T.~R., {Finkbeiner} D.~P., 2010, \apj, 724, 1044

\bibitem[{Tenorio-Tagle} \& {Munoz-Tunon}(1998)]{Tenorio1998MNRAS}
{Tenorio-Tagle} G., {Munoz-Tunon} C., 1998, \mnras, 293, 299

\bibitem[{Thompson} et~al.(2005){Thompson}, {Quataert} \&
  {Murray}]{Thompson2005ApJ}
{Thompson} T.~A., {Quataert} E., {Murray} N., 2005, \apj, 630, 167

\bibitem[{Tombesi} et~al.(2010){Tombesi}, {Cappi}, {Reeves}
  et~al.]{Tombesi2010A&A}
{Tombesi} F., {Cappi} M., {Reeves} J.~N., et~al., 2010, \aap, 521, A57+

\bibitem[{Tremaine} et~al.(2002){Tremaine}, {Gebhardt}, {Bender}
  et~al.]{Tremaine2002ApJ}
{Tremaine} S., {Gebhardt} K., {Bender} R., et~al., 2002, \apj, 574, 740

\bibitem[{Veilleux} et~al.(2001){Veilleux}, {Shopbell} \&
  {Miller}]{Veilleux2001AJ}
{Veilleux} S., {Shopbell} P.~L., {Miller} S.~T., 2001, \aj, 121, 198

\bibitem[{Wagner} et~al.(2012){Wagner}, {Bicknell} \& {Umemura}]{WagnerEtal12}
{Wagner} A.~Y., {Bicknell} G.~V., {Umemura} M., 2012, \apj, 757, 136

\bibitem[{Wagner} et~al.(2013){Wagner}, {Umemura} \& {Bicknell}]{WagnerEtal13}
{Wagner} A.~Y., {Umemura} M., {Bicknell} G.~V., 2013, \apjl, 763, L18

\bibitem[{Wolfire} et~al.(2003){Wolfire}, {McKee}, {Hollenbach} \&
  {Tielens}]{Wolfire2003ApJ}
{Wolfire} M.~G., {McKee} C.~F., {Hollenbach} D., {Tielens} A.~G.~G.~M., 2003,
  \apj, 587, 278

\bibitem[{Zubovas} \& {King}(2012{\natexlab{a}})]{Zubovas2012ApJ}
{Zubovas} K., {King} A., 2012{\natexlab{a}}, \apjl, 745, L34

\bibitem[{Zubovas} \& {King}(2014)]{Zubovas2014arXiv}
{Zubovas} K., {King} A., 2014, ArXiv e-prints

\bibitem[{Zubovas} \& {King}(2012{\natexlab{b}})]{Zubovas2012MNRASb}
{Zubovas} K., {King} A.~R., 2012{\natexlab{b}}, \mnras, 426, 2751

\bibitem[{Zubovas} \& {Nayakshin}(2012)]{Zubovas2012MNRASa}
{Zubovas} K., {Nayakshin} S., 2012, \mnras, 424, 666

\bibitem[{Zubovas} et~al.(2013){Zubovas}, {Nayakshin}, {King} \&
  {Wilkinson}]{Zubovas2013MNRASb}
{Zubovas} K., {Nayakshin} S., {King} A., {Wilkinson} M., 2013, \mnras, 433,
  3079

\end{thebibliography}
\end{document}